\documentclass[journal,10.5pt,onecolumn]{IEEEtran}
\usepackage{amsmath}
\usepackage{geometry}
\usepackage{hyperref}
\usepackage{cite}
\usepackage{sidecap}
\usepackage{bigints}
\usepackage{mathrsfs}
\usepackage{booktabs}
\usepackage[pdftex]{graphicx}
\usepackage{graphics}

\begin{document}

\title{{Outage Analysis in SWIPT Enabled Cooperative AF/DF Relay Assisted Two-Way Spectrum Sharing Communication}}
%
%
%

\author{Sutanu Ghosh, \IEEEmembership{Student Member, IEEE}, Tamaghna Acharya, \IEEEmembership{Member, IEEE}, Santi P. Maity, \IEEEmembership{Member, IEEE}

\thanks{Sutanu Ghosh and Tamaghna Acharya are with the Department of Electronics and Telecommunication Engineering, Indian institute of Engineering Science and Technology, Shibpur, Howrah, West Bengal, 711103, India. E-mail: sutanu99@gmail.com; tamaghna.acharya@ieee.org. \par 
Santi P. Maity is with the Department of Information Technology,  Indian institute of Engineering Science and Technology, Shibpur, Howrah, West Bengal, 711103, India. E-mail: santipmaity@it.iiests.ac.in}}

\maketitle

\begin{abstract}
This paper reports relative performance of decode-and-forward (DF) and amplify-and-forward (AF) relaying in a multi-antenna cooperative cognitive radio network (CCRN) that supports device-to-device (D2D) communications using spectrum sharing technique in cellular network. In this work, cellular system is considered as primary and internet of things devices (IoDs), engaged in D2D communications, are considered to be secondary system. The devices access the licensed spectrum by means of the cooperation in two-way primary communications. Furthermore, IoDs are energized by harvesting the energy from radio frequency (RF) signals, using simultaneous wireless information and power transfer (SWIPT) protocol. Closed form expressions of outage probability for both cellular and D2D communications are derived and the impact of various design parameters for both AF and DF relaying techniques are studied. Based on the simulation results, it is found that the proposed spectrum sharing protocol, for both DF relaying and AF relaying schemes, outperform another similar network architecture in terms of spectrum efficiency. It is also observed that the performance of the proposed system using DF relaying is better than AF relaying scheme in terms of energy efficiency at same transmit power.
\end{abstract}


%

\section{Introduction}
Internet-of-things (IoT) is now under extensive development phase to support the needs of short packet (few information bytes) delivery (through frequency access or sharing) at ultra-high reliability (low outage) and low latency (through efficient routing). Device-to-Device (D2D) communication looks promising in IoT networks due to enhanced battery life and service availability [1]. It also improves the proximity gain and pairing gain of radio spectrum without any involvement of cellular base station (BS), especially when the radio frequency spectrum is shared between D2D and cellular users [2]. 
Congestion due to proliferation of wireless networks in unlicensed band urges the essence of sharing of the spectrum, originally licensed to cellular users, so as to support future IoT communication [3]. 

Cooperative cognitive radio network (CCRN) refers to a network model that facilitates overlay mode of  cognitive radio enabled spectrum sharing [4] and could be applied in D2D communication in 5G  heterogeneous networks (HetNets). Following this, the pair of IoT devices (IoDs), using D2D communication as unlicensed users, may be viewed as secondary users (SUs), while cellular nodes (i.e. evolved Node-B (eNB) and user equipment (UE)) can be modelled as primary users (PUs). IoDs can access the licensed spectrum of cellular system while agreeing to relay the signals of the latter, thereby improving the reliability of  their communication over the fading wireless channel. In literature, two-way communications [5] based on CCRN  is studied to achieve higher spectrum efficiency (SE) over the one-way relaying [6]. The outage performance of the two-way CCRN is studied in [5] and the impact of different relaying schemes on spectrum efficient network operation is reported. 

 Recently, radio frequency energy harvesting (RF-EH) based relaying using simultaneous wireless information and power transfer (SWIPT) protocol has been under investigation to enhance network lifetime and reliability of wireless communication [7]-[9]. SU transmitter of CCRN may follow amplify-and-forward (AF) or, decode-and-forward (DF) relaying to support PU communication and send its own message to SU receiver using the energy harvested from the received PU signal. Relay node harvests the energy from the PU signal by following power splitting (PS) or, time switching (TS) based SWIPT protocols [10].  
Performances of unidirectional or bidirectional communication using AF and DF relay aided SWIPT network are analysed in [11]-[13], [15]. The system outage performances of One-way DF relay assisted communication in CCRN is studied in [11] over a Nakagami fading channel. In [12], PU and SU outage performances are studied in two-way CCRN using AF relay assisted network. The authors of [12] also show the impact of energy conversion efficiency on the system energy efficiency (EE). The performance of similar network model is studied in [13] using DF relay. Based on the study of [13], it is shown that DF relay-assisted two-way communication is significantly more spectrum efficient than one-way communication [11] using PS relaying (PSR) protocol. However, privacy and security issues may be of concern as  DF relay is required to decode PU message during the relaying process [14]. On the contrary, in AF relaying, PU signal does not need to be decoded at the relaying node, although, it is often considered to be an energy inefficient approach as noise gets amplified by the relay node. The performance improvement of SU communication over [13] is studied in [15] using bidirectional SU communication. Needless to mention that the presence of multiple antennas at the source  and or the relay in any SWIPT enabled relay assisted  communication over fading channel would not only improve the reliability of information transfer, but also enhances in energy harvesting at the relay node. Motivated by this, a preliminary study on the performance of the same system model [13] with multiple antenna PUs  is presented in [16]. In this paper, a comparative study between DF and AF relaying scheme is presented using PS protocol in CCRN. 

\subsection*{Scope and Contributions:}
The work in [13] follows three-phase communication using DF relaying with a single-antenna in a CCRN framework. However, it does not explore the impact of multi-antenna on SE and EE aspects of the proposed D2D operation in 5G HetNets. Two-phase communication is more preferable over three-phase to enhance the system throughput. This leads us to explore the problem of spectrum sharing for bidirectional cellular communication and unidirectional D2D communication simultaneously. To offer improved SE, the present cellular system use multiple antennas, the trend seen on 5G [17]. To make this study more general, non-identical number of antennas are considered at two ends of the PU system. The nodes engaged in D2D communication are equipped with necessary hardware to harvest energy from the RF signals of the cellular users. EH meets the energy requirement of the transmitting node for D2D communication to relay (AF/DF) the signals of cellular users and also its own message transmission over the cellular spectrum simultaneously. The main objective of this work is to highlight the relative improvements in SE and EE performances of AF and DF relaying as the PU system moves from single antenna to multi-antenna system. Our contributions can be summarized as follows.

$\bullet$  A novel CCRN architecture with multi-antenna PU system is proposed where two-phase protocol supports two-way SWIPT enabled communications between the pair of cellular users (PUs) and also one-way D2D (SU-to-SU) communication. Two fold benefits, the first one is the improved SE due to multi-antenna and the other one is the throughput improvement due to two-phase protocol are achieved.


$\bullet$ Closed form expressions of both PU and SU outage probability are derived for both DF and AF relay assisted communication using PSR protocol for multi-antenna CCRN framework. Simulation results closely match the analytical expressions.

$\bullet$ The exact dependence of the system performance on various parameters like power sharing factor, power splitting factor, transmission power is also shown through the simulation results. DF relaying mostly performs better than AF relaying. As we have used two-phase spectrum sharing protocol, therefore both of the relaying mechanisms show almost equally spectrum efficiency compared to the similar network architecture [13]. However, AF relaying is found to be better in terms of incremental improvement in SE with  increase in the number of antennas.  

The various symbols used are introduced in Table I. The remaining part of the paper is arranged as follows. The system model is described in Section II and communication protocol description is given in Section III. The outage performance of both PU and SU are analysed in Section IV. The necessary simulation and the numerical results in terms of the various system parameters is presented in Section V. Finally, the paper is concluded in Section VI.
 
 \begin{table}\caption{Symbols and definitions}
  \centering
    \begin{tabular}{l|p{65mm}}
    \hline
    Symbols & Definitions\\
    \hline
    \hline
    $X_s$  & Signal transmission between IoD$_1$ (SU$_1$) and IoD$_2$ (SU$_2$) \\
    \hline
    $P_{p_1}$, $P_{p_2}$  & Transmission power of PU$_1$, PU$_2$\\
    \hline
    $\textbf h_i$  &  Channel gain of link PU$_i$ $\rightarrow$ SU$_1$ ($i \in 1,2$)\\
    \hline
    $\textbf g_i$  &  Channel gain of link PU$_i$ $\rightarrow$ SU$_2$ ($i \in 1,2$)\\
    \hline
    $D_{i}$  &  Distance between PU$_i$ and SU$_1$ ($i \in 1,2$)\\
    \hline
    $D_{j}$  &  Distance between PU$_i$ and SU$_2$ (for i=1, j=4 and i=2, j=5 )\\
    \hline
     $n_{su}$, $n_{pu}$ & Received noise at SU, PU, respectively\\
    \hline 
    $h_{3}$  &  Channel gain of link SU$_1$ $\rightarrow$ SU$_2$  \\
    \hline
    $\rho$  &  Power splitting factor of SU$_1$\\
    \hline
    $\alpha$ &  Power sharing factor at SU$_1$\\
    \hline
     L  &  Distance between PU$_1$ and PU$_2$\\
    \hline
    $D_3$  &   Distance between SU$_1$ and SU$_2$\\
    \hline
    $m$ &  Path loss exponent\\
    \hline
    $\eta$  &  Energy conversion efficiency at SU$_1$ \\
    \hline
    $R_{PU}$, $R_{SU}$  &  Target rate of PU and SU communication, respectively \\
    \hline
    \hline
    \end{tabular}
  \end{table}

\section{System model}
\subsection{Assumptions and Notations}
The system model consists of a pair of eNB and UE in long term evolution (LTE) network architecture as depicted in Fig. 1. The eNB and UE are equipped with multiple antennas N$_a$ and N$_b$, respectively. In absence of direct communication link, the cellular nodes (eNB and UE) intend to exchange their information via a single antenna IoT device (IoD) (denoted as IoD$_1$ in Fig. 1.), aiming to achieve the target rate of $R_{PU}$ at each side. Simultaneously, IoD$_1$ sends its own message signal to IoD$_2$ to meet a target rate of $R_{SU}$. Both of them use single antenna for communication. Here eNB and UE are considered as PU$_1$ and PU$_2$, respectively and IoD$_1$ and IoD$_2$ are modelled as SU$_1$ and SU$_2$, respectively. \par

\begin{figure}[h]
  \centering
  \includegraphics[width=0.78\linewidth]{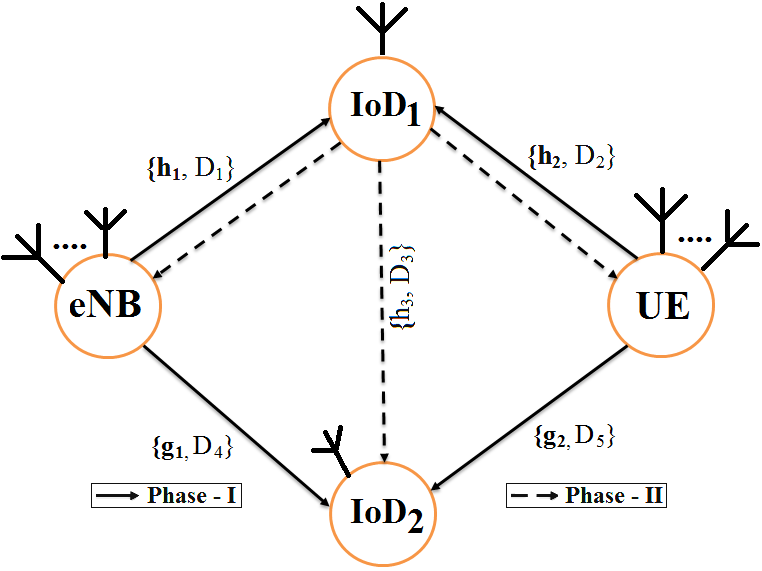}
  \caption{\textbf{System model}}
  \label{f12}
\end{figure}

We assume that PU$_i$, ($i \in 1,2$) uses fixed power supply, i.e., $P_{p_i}$, SU$_1$ is powered through harvested energy from PUs' RF signals ,using SWIPT, for relaying PUs' messages and transmitting to SU$_2$. We consider both DF and AF relaying mechanisms to support the two-way PU communications. $\textbf h_i= [h_{i,1}, h_{i,2}, . . . , h_{i,N_p}]$, (N$_p$ $\in$ N$_a$, N$_b$) and $\textbf g_i= [g_{i,1}, g_{i,2}, . . . , g_{i,N_p}]$, (i $\in$ 1,2) are the vector channel coefficients from the multiple-antenna PU$_i$ to SU$_1$ and PU$_i$ to SU$_2$, respectively, where $h_{i,n} \sim\mathcal{CN}(0,1)$ and $g_{i,n} \sim\mathcal{CN}(0,1)$. Channel between $SU_1 \rightarrow SU_2$ is represented by $ h_3$, where $h_3 \sim\mathcal{CN}(0,1)$. All the channel distribution between PU and SU links follow independent and identically distributed (i.i.d) Rayleigh fading. Due to short distance the link between SU$_1$ and SU$_2$ is considered as Nakagami-m fading. The instantaneous channel state information (CSI) is assumed to be unavailable at PU$_i$ ($i$ $\in$ 1, 2). It is also considered that the full-diversity space-time codes (like GABBA codes [18]) are used at the PU nodes. Normally, in space-time code, power is uniformly distributed among the transmitting antennas. The distances between the users PU$_1-$SU$_1$, PU$_2-$SU$_1$, PU$_1-$SU$_2$, PU$_2-$SU$_2$ and SU$_1-$SU$_2$ are given by $D_1$, $D_2$, $D_4$, $D_5$  and $D_3$, respectively with \textquoteleft\textit{m}\textquoteright  as the path-loss exponent. Here $n_{su}$ and $n_{pu}$ indicate the received noise at SU and PU, respectively. These noises are additive white Gaussian noise (AWGN) with zero (0) mean and the variance $\sigma ^2$. 
\subsection{Protocol description}
\begin{figure}[h]
  \centering
  \includegraphics[width=0.78\linewidth]{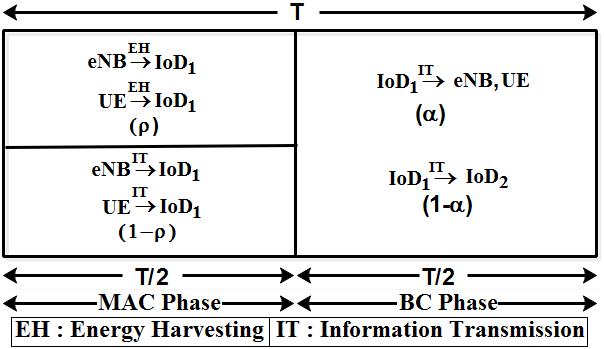}
  \caption{\textbf{Transmission frame structure in the two-way PSR protocol}}
  \label{f13}
\end{figure} 

The two-way communications between PU$_1$ and PU$_2$ via SU$_1$ take place in two time phases viz., multiple access channel (MAC) and broadcast (BC) phase, as shown in Fig. 2. In the MAC phase, both PU$_1$ and PU$_2$ simultaneously transmit their information signals to SU$_1$. SU$_1$ is able to harvest energy from part of the received signal using PSR protocol. In the case of AF relaying, SU$_1$ broadcasts an amplified PU signal superimposed with the SU signal $X_s$ in the BC phase, using the total harvested energy. The PU receivers are able to receive the signals using the maximal-ratio-combining (MRC) and separate the desired signal using the self-interference cancellation (SIC). Now, decoding of SU$_1$'s message at SU$_2$ is performed in the two phases; first the stronger PU signal is decoded considering SU$_1$'s desired signal as interference. Once the strong PU signal is separated from the received signal in the the first phase, SU$_2$ decodes the desired SU$_1$'s signal in presence of channel noise. 

In the case of DF relaying, SU$_1$ first decodes the PUs' message signals received in the MAC phase, and then broadcasts a network coded primary signal superimposed on the secondary signal $X_s$ in the BC phase. Both the PU nodes decode the network coded (XOR operation) PU signal in presence of SU interfering signal and noise. Thereafter, $SU_{2} $ decodes $ SU_{1} $ signal by cancelling PU signal's based on the PUs' messages received previously in the MAC Phase.

\section{Rate and Outage Analysis in DF relaying}
\subsection{\textbf {Rate Analysis}}
Now the received power in MAC phase at ${SU_1}$ from PU$_1$ and PU$_2$ can be expressed as follows\footnote{{In absence of CSI, PU$_1$ and PU$_2$ are assumed to use suitable space-time coding techniques, the details of which are not within the scope of our current study.}} [19]. 
\begin{align} 
\scalebox{01}{$P_{SU_1}^{(1)}={\frac{P_{p_1}}{N_a (D_1)^{m}}} \sum_{n=1}^{N_a} \mid{h}_{1,n}\mid^2  + {\frac{P_{p_2}}{N_b (D_2)^{m}}} \sum_{p=1}^{N_b} \mid{h}_{2,p}\mid^2 $}
\end{align}

where $P_{p_1}$ and $P_{p_2}$ are the transmit powers of PU$_1$ and PU$_2$, respectively.

Similarly, the received power at ${SU_2}$ from PU$_1$ and PU$_2$ can be expressed as
\begin{align} \label{q1}
\scalebox{01}{$P_{SU_2}^{(1)}={\frac{P_{p_1}}{N_a (D_4)^{m}}} \sum_{n=1}^{N_a} \mid{g}_{1,n}\mid^2  + {\frac{P_{p_2}}{N_b (D_5)^{m}}} \sum_{p=1}^{N_b} \mid{g}_{2,p}\mid^2   $}
\end{align}

A part $\rho$ (referred to as power splitting factor in the subsequent discussion) of the received signal  power at SU$_1$ is used for energy harvesting, and the rest (1-$\rho$) portion is used for information processing. The harvested energy in MAC phase is given by 
\begin{align} \label{q2}
\scalebox{01}{$E_s = \eta \rho \big(\frac{P_{p_1}}{N_a (D_1)^m} \sum_{n=1}^{N_a}\mid{h}_{1,n}\mid ^2 + \frac{P_{p_2}}{N_b (D_2)^m} \sum_{p=1}^{N_b}\mid{h}_{2,p}\mid ^2 \big) \frac{T}{2}$}
\end{align}

where 0 $< \eta <$ 1 represents the energy conversion efficiency. After power splitting, the received power in MAC  phase at the receiver of SU$_1$ for information processing is given by
\begin{align} \label{q3}
\scalebox{01}{$P_{SU_1}^{(1,IT)}=(1-\rho)\big[{\frac{P_{p_1}}{N_a (D_1)^m}} \sum_{n=1}^{N_a} \mid{h}_{1,n}\mid^2 + {\frac{P_{p_2}}{N_b (D_2)^m}} \sum_{p=1}^{N_b} \mid{h}_{2,p}\mid^2\big] $}
\end{align}

Now following the linear EH model, the available power at SU$_1$ for transmission in the BC phase is given by
\begin{align} \label{q2}
\scalebox{1}{$P_s = \eta \rho \big(\frac{P_{p_1}}{N_a (D_1)^m} \sum_{n=1}^{N_a}\mid{h}_{1,n}\mid ^2 + \frac{P_{p_2}}{N_b (D_2)^m} \sum_{p=1}^{N_b}\mid{h}_{2,p}\mid ^2 \big)$}
\end{align}

In the MAC phase, the successful decoding of the received signals from PU$_1$ and PU$_2$ is possible at SU$_{1}$, if $R_{SU_{1}}^{(1)} \geq R_{PU}$, $R_{SU_{1}}^{(2)} \geq R_{PU}$ and $R_{SU_{1}}^{\sum} \geq 2R_{PU}$ [20], where
\begin{align}
Q_1=\begin{cases} 
\scalebox{01}{$R_{SU_{1}}^{(11)} = \frac{T}{2T}\log_2 \big(1+\frac{(1-\rho) P_{p_1}}{N_a (D_{1})^m \sigma^2} \sum_{n=1}^{N_a}\mid{h}_{1,n}\mid ^2 \big)$}, \\
\scalebox{01}{$R_{SU_{1}}^{(12)} = \frac{T}{2T}\log_2 \big(1+\frac{(1-\rho) P_{p_2}}{N_b (D_{2})^m \sigma^2} \sum_{p=1}^{N_b}\mid{h}_{2,p}\mid ^2 \big)$},  \\
\scalebox{01}{$R_{SU_{1}}^{\sum} = \frac{T}{2T}\log_2 \big(1+\frac{(1-\rho) P_{p1}}{N_a(D_{1})^m \sigma^2} \sum_{n=1}^{N_a}\mid{h}_{1,n}\mid ^2 + \frac{(1-\rho) P_{p2}}{N_b(D_{2})^m \sigma^2} \sum_{p=1}^{N_b}\mid{h}_{2,p}\mid ^2\big)$} 
\end{cases}
\end{align}

Similarly, in the MAC phase, the successful decoding of the received signals from PU$_1$ and PU$_2$ is possible at SU$_{2}$, if $R_{SU_{2}}^{(1)} \geq R_{PU}$, $R_{SU_{2}}^{(2)} \geq R_{PU}$ and $R_{SU_{2}}^{\sum} \geq 2R_{PU}$, where
\begin{align} 
 Q_2=\begin{cases}
\scalebox{01}{$R_{SU_{2}}^{(11)} = \frac{1}{2}\log_2 \big(1+\frac{P_{p_1}}{N_a (D_{4})^m \sigma^2} \sum_{n=1}^{N_a}\mid{g}_{1,n}\mid ^2 \big)$}, \\
\scalebox{01}{$R_{SU_{2}}^{(12)} = \frac{1}{2}\log_2 \big(1+\frac{P_{p_2}}{N_b (D_{5})^m \sigma^2} \sum_{p=1}^{N_b}\mid{g}_{2,p}\mid ^2 \big)$},  \\
\scalebox{01}{$R_{SU_{2}}^{\sum} = \frac{1}{2}\log_2 \big(1+\frac{ P_{p1}}{N_a(D_{4})^m \sigma^2} \sum_{n=1}^{N_a}\mid{g}_{1,n}\mid ^2 + \frac{P_{p2}}{N_b(D_{5})^m \sigma^2} \sum_{p=1}^{N_b}\mid{g}_{2,p}\mid ^2\big)$} 
\end{cases}
\end{align}

In DF relaying, SU$_1$ uses $\alpha$ ($0<\alpha<1$) fraction of its total transmit power $P_s$ to relay the network-coded primary information and rest (1-$\alpha$) fraction of $P_s$ is used to send its own independent message $X_s$ to SU$_2$.


After MRC and cancellation of self-interference terms by applying SIC technique, the received signal-to-interference-noise ratio (SINR) at PU$_i$ (i $\in$ 1,2; N$_p$ $\in$ N$_a$, N$_b$) receivers in BC phase can be expressed as
\begin{align} 
\scalebox{1}{$\gamma_{i}^{DF} = \frac{\frac{\alpha P_s}{(D_{i})^m} \sum_{w=1}^{N_p}{\mid {h}_{i,w} \mid}^2}{\frac{(1-\alpha)P_s}{(D_{i})^m} \sum_{w=1}^{N_p}{\mid h_{i,w} \mid}^2 + \sigma^2}$}
\end{align}

It is assumed that SU$_2$, like SU$_1$, succeeds in decoding the PUs' signals received in MAC phase. Based on these prior knowledges, SU$_2$ is able to separate the desired SU signal from the PUs' interference received in BC phase [13]. Therefore, the received signal at $SU_2$ can be rewritten as
\begin{align} 
\scalebox{01}{$Y_{SU_{2}}^{(2,DF)} = \underbrace{\sqrt{\frac{(1-\alpha)P_s}{{D_3}^m}}{h}_3 X_s}_{\textbf{Required signal}}+ \underbrace{n_{su}}_{\textbf{Noise}}$}
\end{align}

Now, the achievable rate at $PU_i$ is given by (based on the SINR at $PU_i$) ($w$ $\in$ $n,p$)
\begin{align} 
\scalebox{01}{$R_{PU_i}^{(2,DF)}= \frac{1}{2} \log_2 \lbrace 1+\gamma_{i}^{DF}\rbrace= \frac{1}{2} \log_2 \bigg\lbrace 1+\frac{{a_{i}^\prime} (\sum_{n=1}^{N_a} a_1{\mid {h}_{1,n} \mid}^2 + \sum_{p=1}^{N_b} b_1{\mid {h}_{2,p} \mid}^2) \sum_{w=1}^{N_p} {\mid {h}_{i,w} \mid}^2}{{b_{i}^\prime} (\sum_{n=1}^{N_a} a_1{\mid {h}_{1,n} \mid}^2 +\sum_{p=1}^{N_b} b_1{\mid {h}_{2,p} \mid}^2) \sum_{w=1}^{N_p}{\mid {h}_{i,w} \mid}^2 + 1}\bigg\rbrace$}
\end{align}

where $\scalebox{01}{$a_1={\eta}\rho\frac{P_{p_1}}{N_a (D_{1})^m   \sigma^2}$}$, $\scalebox{01}{$b_1={\eta}\rho\frac{P_{p_2}}{N_b (D_{2})^m  \sigma^2}$}$, $\scalebox{01}{$a_{i}^\prime=\dfrac{\alpha}{(D_{i})^m  }$}$, $\scalebox{01}{$b_{i}^\prime=\dfrac{(1-\alpha)}{(D_{i})^m   }$}$.

Based on (9), the achievable rate at $SU_2$ is expressed as
\begin{align}
\scalebox{01}{$R_{SU_{2}}^{(2,DF)}= \frac{1}{2} \log_2 \big\lbrace 1+\frac{\frac{(1-\alpha) P_s}{{D_3}^m} {\mid {h}_3 \mid}^2}{\sigma^2}\big\rbrace = \frac{1}{2} \log_2 \big\lbrace 1+{{c} (\sum_{n=1}^{N_a} a_1{\mid {h}_{1,n} \mid}^2 +\sum_{p=1}^{N_b} b_1{\mid {h}_{2,p} \mid}^2) }{\mid {h}_3 \mid}^2\big\rbrace$}
\end{align}

where $\scalebox{01}{$c=\dfrac{(1-\alpha)}{{D_3}^m }$}$.

\subsection{\textbf {Outage Probability Analysis}}
An outage occurs when the achievable rate of data transmission on any transmission link falls below the target rate of data transmission. 
\subsubsection{Outage Probability Analysis of Primary System}
 Based on the definition, PU outage probability\footnote{Since SU$_1$ is used as a relay to support PU communication, therefore the successful information transmission from both the PU nodes to SU$_1$ is essential in MAC phase for relaying the information in BC phase.} using DF relaying mechanism can be determined as follows [13]:
\begin{multline}
\hspace*{3.3cm}\mathscr{P}_{out}^{(PU,DF)} = 1 - \overbrace{\bigg[\underbrace{\mathscr{P}\lbrace Q_1 \rbrace}_{\textbf{MAC phase}} \times  \underbrace{\mathscr{P}\big\lbrace min(R_{PU_1}^{(2,DF)},R_{PU_2}^{(2,DF)})\geq R_{PU}\big\rbrace}_{\textbf{BC phase}}\bigg]}^{\textbf{Succes Probability}}\\=1 - \bigg[\mathscr{P}\lbrace Q_1 \rbrace\times  \mathscr{P}\big\lbrace R_{PU_1}^{(2,DF)}\geq R_{PU}\big\rbrace \times \mathscr{P}\big\lbrace R_{PU_2}^{(2,DF)}\geq R_{PU}\big\rbrace\bigg]
\end{multline}
Applying (6), the success probability of data transmission between both of the PU nodes (PU$_1$ and PU$_2$) and SU$_1$ is expressed by following three conditions satisfied together.
{{\begin{align}
 Q_1=\begin{cases}
\scalebox{01}{$R_{SU_{1}}^{(11)} = {\frac{1}{2} }\log_2 \big(1+A_1 {X}_{1} \big)\geq R_{PU}$},\\
\scalebox{01}{$R_{SU_{1}}^{(12)} = {\frac{1}{2}}\log_2 \big(1+A_2 {Y}_{1} \big)\geq R_{PU}$},\\
\scalebox{01}{$   R_{SU_{1}}^{\sum}= {\frac{1}{2}}\log_2 \big(1+ A_1 {X}_{1} + A_2 {Y}_{1} \big)\geq 2R_{PU}$}
\end{cases}
\end{align}}}
where $\scalebox{01}{$A_1=\frac{(1-\rho) P_{p_1}}{N_a (D_{1})^m \sigma^2}$}$, $\scalebox{01}{$A_2=\frac{(1-\rho) P_{p_2}}{N_b (D_{2})^m \sigma^2},$}$, $\scalebox{01}{$u_{1}=2^{\big(2R_{PU}\big)}-1$}$, $\scalebox{01}{$u_{2}=2^{\big(2R_{PU}\big)}-1$}$, $\scalebox{01}{$u_{3}=2^{\big(4R_{PU}\big)}-1$}$.

$X_1=\sum_{n=1}^{N_a}\mid{h}_{1,n}\mid ^2$ and $Y_1=\sum_{p=1}^{N_b}\mid{h}_{2,p}\mid ^2$ follow the same nature of gamma distribution with $N_p$ degree of freedom. The probability density function (PDF) can be expressed as $\scalebox{0.92}{$f_{X}{(x)}=\frac{x^{N_p-1} \text{exp}\big(-\frac{x}{1/N_p}\big)}{\Gamma(N_p) \big(1/N_p\big)^{N_p}}$}$. Based on (13), $\mathscr{P}\lbrace Q_1 \rbrace$ can be expressed as [21, Sec. 3.381.1, 3.381.3, 8.350.1, 8.352.1]
\begin{multline} 
\scalebox{0.8}{$\mathscr{P}\lbrace Q_1 \rbrace=\int\limits_{\frac{u_{1}}{A_1}}^\infty \frac{{x_1}^{N_a-1} \text{exp}\big(-\frac{x_1}{1/N_a}\big)}{\Gamma(N_a) \big(1/N_a\big)^{N_a}} \Bigg\{\int\limits_{\frac{u_{2}}{A_2}}^\infty \frac{{y_1}^{N_b-1} \text{exp}\big(-\frac{y_1}{1/N_b}\big)}{\Gamma(N_b) \big(1/N_b\big)^{N_b}}dy_{1}\Bigg\}dx_{1}-\int\limits_{\frac{u_{1}}{A_1}}^{\frac{u_{3}-u_{1}}{A_1}} \frac{{x_1}^{N_a-1} \text{exp}\big(-\frac{x_1}{1/N_a}\big)}{\Gamma(N_a) \big(1/N_a\big)^{N_a}} \Bigg\{\int\limits_{\frac{u_{2}}{A_2}}^{\frac{u_{3}-A_1 x_{1}}{A_2}} \frac{{y_1}^{N_b-1} \text{exp}\big(-\frac{y_1}{1/N_b}\big)}{\Gamma(N_b) \big(1/N_b\big)^{N_b}}dy_{1}\Bigg\}dx_{1}$}\\\scalebox{0.9}{$=\frac{1}{\Gamma(N_a)\Gamma(N_b)}{\Gamma\Bigg(N_a,{{N_a u_1}/{A_1}}\Bigg)\Gamma\Bigg(N_b,{{N_b u_2}/{A_2}}\Bigg)}-\Bigg\{\frac{1}{\Gamma({N_a})}-\frac{\gamma\Big(N_b,N_b \frac{u_{2}}{A_2}\Big)}{\Gamma(N_a)\Gamma(N_b)}\Bigg\}\Big\{\gamma\big(N_a,N_a\frac{u_{3}-u_{1}}{A_1}\big)-\gamma\big(N_a,N_a\frac{u_1}{A_1}\big)\Big\}$}\\\scalebox{0.9}{$+\sum_{q_a=0}^{N_b-1}\frac{\exp\big(-\frac{N_b {u_{3}}}{A_2}\big)\sum_{q=0}^{q_a}(-1)^{q}{{q_a}\choose{q}} {u_3}^{{q_a}-q} }{{q_a}!  \Gamma(N_a) (1/N_a)^{N_a} \big(1/N_b\big)^{q_a} {A_2}^{q_a}}\frac{{A_1}^q}{\Big(N_a-\frac{A_1 N_b}{A_2}\Big)^{(N_a+q)}}\Big\{\gamma\big[N_a+q, \big(N_a-\frac{A_1 N_b}{A_2}\big)\frac{u_{3}-u_{1}}{A_1}\big]-\gamma\big[N_a+q, \big(N_a-\frac{A_1 N_b}{A_2}\big)\frac{u_{1}}{A_1}\big]\Big\}$}
\end{multline}

\textbf{Proof} : See Appendix A.

where $\Gamma(.,.)$ and $\gamma(.,.)$ are the upper and the lower incomplete gamma function, respectively.

The probability of successful data transmission from SU$_1$ to PU$_1$ can be written as
\begin{align} 
\scalebox{01}{$R_{PU_1}^{(2,DF)}= \frac{1}{2} \log_2 \lbrace 1+\gamma_{1}^{DF}\rbrace= \frac{1}{2} \log_2 \Bigg\lbrace 1+\frac{{a_{1}^\prime}  (\sum_{n=1}^{N_a} a_1{\mid {h}_{1,n} \mid}^2 + \sum_{p=1}^{N_b} b_1{\mid {h}_{2,p} \mid}^2) \sum_{n=1}^{N_a}{\mid {h}_{1,n} \mid}^2}{{b_{1}^\prime} (\sum_{n=1}^{N_a} a_1{\mid {h}_{1,n} \mid}^2 +\sum_{p=1}^{N_b} b_1{\mid {h}_{2,p} \mid}^2)\sum_{n=1}^{N_a} {\mid {h}_{1,n} \mid}^2 + 1}\Bigg\rbrace$}
\end{align}
\small{\begin{multline} 
\scalebox{01}{$\mathscr{P}\Bigg\{R_{PU_1}^{(2,DF)}\geq R_{PU}\Bigg\}=\mathscr{P}\Bigg\{\frac{{a_{1}^\prime} (a_1 X_1+b_1 Y_1)X_1 }{{b_{1}^\prime} (a_1 X_1+b_1 Y_1)X_1 + 1} \geq u_1 \Bigg\}$}= \begin{cases}
\scalebox{01}{$1 - \mathscr{P}\Bigg\{X_1 \leq \frac{k^{'}}{(a_1 X_1+b_1 Y_1)} \Bigg\},\hspace*{0.85cm}{\text for,} \hspace*{0.15cm} u_1 < \frac{\alpha}{(1-\alpha)}$}\\
\scalebox{01}{$1 - \mathscr{P}\Bigg\{X_1 \geq \frac{k^{'}}{(a_1 X_1+b_1 Y_1)} \Bigg\} = 0,$} \hspace*{0.4cm} otherwise .\\
\end{cases}
\end{multline}}
where $k^{'} = \frac{{u_1}}{a_{1}^\prime - {u_1} b_{1}^\prime}$.
Now, the solution to (16) is obtained as (17) [21, Sec. 8.350.1, 8.352.1].
\begin{multline} 
\scalebox{0.95}{$\mathscr{P}\Bigg\{X_1 \leq \frac{k^{'}}{(a_1 X_1+b_1 Y_1)} \Bigg\}=\mathscr{P}\Bigg( Y_1\leq \frac{k^{'}}{b_{1} X_{1}}-\frac{a_{1} X_{1}}{b_{1}}\Bigg)
=\frac{1}{\Gamma(N_a)}{\gamma\Bigg(N_a,\frac{\sqrt{\frac{k^{'} }{a_{1}}}}{1/N_a}\Bigg)}-\Bigg[\sum_{p_a=0}^{N_b-1}\frac{{N_a}^{N_a} {N_b}^{p_a}}{p_a!  \Gamma(N_a)}\sum_{r=0}^{p_a}(-1)^{r}{{p_a}\choose{r}} $}\\ \scalebox{0.95}{$ \bigg(\frac{k^{'}}{b_{1}}\bigg)^{p_a-r} \bigg(\frac{a_{1}}{b_{1}}\bigg)^{r} \sum_{l=0}^{\infty} \sum_{t_l=0}^{l} {{l}\choose{t_l}} \frac{(-1)^l}{l! (2r+ N_a -p_a-l+2t_l)}  {\Big(\frac{k^{'} N_b}{b_{1}}\Big)}^{l-t_l} \Bigg({N_a-\frac{a_{1}N_b}{b_{1}}} \Bigg)^{t_l}  \Bigg(\sqrt{\frac{k^{'}}{a_{1}}}\Bigg)^{2r+ N_a -p_a-l+2t_l} \Bigg]$}
\end{multline}

\textbf{Proof} : See Appendix B.
 
Similarly, the probability of successful data transmission from SU$_1$ to PU$_2$ can be written as
\begin{multline} 
\scalebox{0.95}{$\mathscr{P}\Bigg\{Y_1 \leq \frac{k^{''}}{(a_1 X_1+b_1 Y_1)} \Bigg\}=\mathscr{P}\Bigg( X_1\leq \frac{k^{''}}{a_{1} Y_{1}}-\frac{b_{1} Y_{1}}{a_{1}}\Bigg)
=\frac{1}{\Gamma(N_b)}{\gamma\Bigg(N_b,\frac{\sqrt{\frac{k^{''} }{b_{1}}}}{1/N_b}\Bigg)}-\Bigg[\sum_{p_a=0}^{N_a-1}\frac{{N_b}^{N_b}{N_a}^{p_a}}{p_a!  \Gamma(N_b)}\sum_{r=0}^{p_a}(-1)^{r}{{p_a}\choose{r}}$}\\ \scalebox{0.95}{$ \bigg(\frac{k^{''}}{a_{1}}\bigg)^{p_a-r} \bigg(\frac{b_{1}}{a_{1}}\bigg)^{r} \sum_{l=0}^{\infty} \sum_{t_l=0}^{l} {{l}\choose{t_l}} \frac{(-1)^l}{l! (2r+ N_b -p_a-l+2t_l)}  {\Big(\frac{k^{''} N_a}{a_{1}}\Big)}^{l-t_l} \Bigg({N_b-\frac{b_{1}N_a}{a_{1}}}\Bigg)^{t_l}  \Bigg(\sqrt{\frac{k^{''}}{b_{1}}}\Bigg)^{2r+ N_b -p_a-l+2t_l} \Bigg]$}
\end{multline}

where $k^{''} = \frac{{u_1}}{a_{2}^\prime - {u_1} b_{2}^\prime}$.

The closed form solution to PU outage probability can be determined using (14), (17)-(18).

Closed form expression of PU outage probability using DF relaying is given in Appendix F.

\subsubsection{Outage Probability Analysis of Secondary System}
SU outage probability\footnote{Since SU$_2$ needs to decode the message of SU$_1$ by removing PUs message in BC phase, therefore successful information transmission from both PUs to SU$_1$ is necessary in MAC phase.} using DF relaying is shown as follows:
\begin{align} 
\scalebox{1}{$\mathscr{P}_{out}^{(SU,DF)} = 1 - \overbrace{\bigg[\underbrace{\mathscr{P}\lbrace Q_1  \rbrace \times \mathscr{P}\lbrace  Q_2 \rbrace}_{\textbf{MAC Phase}} \times \underbrace{\mathscr{P}\big\lbrace R_{SU_2}^{(2,DF)}\geq R_{SU}\big\rbrace}_{\textbf{BC Phase}}\bigg]}^{\textbf{Success Probability}}$}
\end{align}

Similar to (13), the probability of success due to data transmission from PU$_i$ ($i \in 1,2$) to SU$_2$ is defined by the following three conditions together. 
{\small{\begin{align}
 Q_2= \begin{cases}
\scalebox{01}{$R_{SU_{2}}^{(11)} = {\frac{1}{2} }\log_2 \big(1+B_1 {X}_{2} \big)\geq R_{PU}$},\\
\scalebox{01}{$R_{SU_{2}}^{(12)} = {\frac{1}{2}}\log_2 \big(1+B_2 {Y}_{2} \big)\geq R_{PU}$},\\
\scalebox{01}{$   R_{SU_{2}}^{\sum}= {\frac{1}{2}}\log_2 \big(1+ B_1 {X}_{2} + B_2 {Y}_{2} \big)\geq 2R_{PU}$}
\end{cases}
\end{align}}}
where $\scalebox{01}{$B_1=\frac{P_{p_1}}{N_a (D_{1})^m \sigma^2}$}$, $\scalebox{01}{$B_2=\frac{P_{p_2}}{N_b (D_{2})^m \sigma^2},$}$.

$X_2=\sum_{n=1}^{N_a}\mid{g}_{1,n}\mid ^2$ and $Y_2=\sum_{p=1}^{N_b}\mid{g}_{2,p}\mid ^2$ follow the same nature of gamma distribution like X$_1$ and Y$_1$, respectively.

Based on (20), $\mathscr{P}\lbrace Q_2 \rbrace$ can be expressed as follows
\begin{multline} 
\scalebox{0.9}{$\mathscr{P}\lbrace Q_2 \rbrace=\frac{1}{\Gamma(N_a)\Gamma(N_b)}{\Gamma\Bigg(N_a,{{N_a u_1}/{B_1}}\Bigg)\Gamma\Bigg(N_b,{{N_b u_2}/{B_2}}\Bigg)}-\Bigg\{\frac{1}{\Gamma(N_a)}-\frac{\gamma\Big(N_b,N_b \frac{u_{2}}{B_2}\Big)}{\Gamma(N_a)\Gamma(N_b)}\Bigg\} \Big\{\gamma\big(N_a,N_a\frac{u_{3}-u_{1}}{B_1}\big)-\gamma\big(N_a,N_a\frac{u_1}{B_1}\big)\Big\}$}\\\scalebox{0.9}{$ +\sum_{{q_a}=0}^{N_b-1}\frac{\exp\big(-\frac{N_b {u_{3}}}{B_2}\big)\sum_{q=0}^{{q_a}}(-1)^{q}{{{q_a}}\choose{q}} {u_3}^{{q_a}-q} }{{q_a}!  \Gamma(N_a) \big(1/N_a\big)^{N_a} \big(1/N_b\big)^{{q_a}}{B_2}^{q_a}} \frac{{B_1}^q}{\Big(N_a-\frac{B_1 N_b}{B_2}\Big)^{(N_a+q)}}\Big\{\gamma\big[N_a+q, \big(N_a-\frac{B_1 N_b}{B_2}\big)\frac{u_{3}-u_{1}}{B_1}\big]-\gamma\big[N_a+q, \big(N_a-\frac{B_1 N_b}{B_2}\big)\frac{u_{1}}{B_1}\big]\Big\}$}
\end{multline}

Following (11), the probability of successful data transmission between the link SU$_1$ to SU$_2$ is expressed as 
\begin{align} 
\begin{split}
\scalebox{01}{$\mathscr{P}( R_{SU_2}^{(2,DF)}\geq R_{SU})=\mathscr{P}\Big( Z\geq \frac{u_4}{c(a_1 X_1 +b_1 Y_1)}\Big)$}
\end{split}
\end{align}
where $u_4 = 2^{(2R_{SU})} -1$. $Z = \mid h_3 \mid^2$ and it follows the gamma distribution with
Nakagami shaping parameter m$_k$. The PDF of $Z$ can be described
as $\scalebox{01}{$f_{Z}{(z)}=\frac{z^{m_k-1} \text{exp}\big(-\frac{z}{1/m_k}\big)}{\Gamma(m_k) \big(1/m_k\big)^{m_k}}$}$.

Now (22) can be evaluated as follows [21, Sec. 3.471.9, 8.352.2]:
\begin{multline} 
\scalebox{.9}{$\mathscr{P}( R_{SU_2}^{(2,DF)}\geq R_{SU}) =  \sum_{p_a=0}^{N_b-1}  \frac{\Big(\frac{N_b}{b_1}\Big)^{p_a}}{p_a!  ({1/N_a})^{N_a} \Gamma(N_a)} \sum_{r=0}^{p_a}(-1)^{r}{{p_a}\choose{r}}  \bigg(\frac{u_4}{c}\bigg)^{p_a-r} \big({a_{1}}\big)^{r}\frac{1}{{\bigg(N_a-\frac{a_{1} N_b}{b_{1}}\bigg)}^{N_a+r}}\frac{\Gamma{(N_a +r)}}{\{({1/m_k})^{ m_k}\} \Gamma(m_k)}  $}\\\scalebox{0.9}{$\Bigg[2\bigg\{\frac{u_4 N_b}{c b_{1} m_k} \bigg\}^{(m_k-p_a+r)/2}  K_{m_k-p_a+r}\bigg\{2\sqrt{\frac{u_4 {N_b}m_k}{c b_{1}}}\bigg\}  - \sum_{j=0}^{N_a+r-1} \frac{1}{j!} \bigg[\frac{u_4  \bigg({N_a}-\frac{a_{1}{N_b}}{b_{1}}\bigg)}{c a_{1}} \bigg]^{j} 2\bigg\{\frac{u_4 \bigg(N_a -\frac{a_{1} N_b}{b_{1}}\bigg)}{c a_{1} m_k}+\frac{u_4 N_b}{c b_{1} m_k}\bigg\}^{\frac{(m_k-p_a+r-j)}{2}} $}\\\scalebox{0.9}{$K_{m_k-p_a+r-j}\bigg\{2\sqrt{\frac{u_4 m_k \bigg({N_a}-\frac{a_{1}{N_b}}{b_{1}}\bigg)}{c a_{1}}+\frac{u_4 {N_b}m_k}{c b_{1} }}\bigg\}  \Bigg]+\Bigg[\sum_{p_a=0}^{N_a-1}\frac{\bigg({\frac{u_{4} N_a}{c a_{1}}}\bigg)^{p_a}  ({m_k})^{m_k}}{p_a! \Gamma(m_k)} 2  {\Bigg(      {\frac{u_{4} N_a}{c a_{1} m_k}}\Bigg)}^{(m_k - p_a)/2}K_{m_k - p_a}\bigg(2\sqrt{{{\frac{u_{4} {N_{a}}m_k}{c a_{1} }}}}\bigg)\Bigg]$}
\end{multline}

\textbf{Proof} : See Appendix C.
  
Finally, the closed form SU outage expression using DF relaying can be determined using (14), (21) and (23).

 Closed form expression of SU outage probability using DF relaying is given in Appendix G.

 \section{Rate \& Outage Analysis of AF relaying}
 \subsection{\textbf {Rate Analysis}}
 In the BC phase of AF relaying, SU$_1$ broadcasts an amplified version of the signal generated after combining the PU signals, received in the MAC phase. To include this in the analysis, harvested power P$_s$ in (5) is normalized using the factor $\xi$\footnote{{As noise power is negligible with respect to the power received for information processing, therefore noise power is neglected.}}, expressed as
\begin{multline} \label{q2}
\scalebox{01}{$\xi = \sqrt{\frac{1}{(1-\rho) \bigg[\frac{P_{p_1}}{N_a (D_1)^m} \sum_{n=1}^{N_a}\mid{h}_{1,n}\mid ^2 + \frac{P_{p_2}}{N_b (D_2)^m} \sum_{p=1}^{N_b}\mid{h}_{2,p}\mid ^2 \bigg] +\sigma^2}}$}\\\scalebox{01}{$ \approx \sqrt{\frac{1}{(1-\rho) \bigg[\frac{P_{p_1}}{N_a (D_1)^m} \sum_{n=1}^{N_a}\mid{h}_{1,n}\mid ^2 + \frac{P_{p_2}}{N_b (D_2)^m} \sum_{p=1}^{N_b}\mid{h}_{2,p}\mid ^2 \bigg]}}$}
\end{multline}

 Based on the superposition coding and AF relaying principle SU$_1$ uses $\alpha$ ($0<\alpha<1$) the fraction of its total transmit power $P_s$ to relay the combined signal of primary information and the rest (1-$\alpha$) portion is used to send its own independent message $X_s$ to SU$_2$. 
 As both PUs know their individual transmitted signals, consequently, they are able to cancel their self-interference terms. Therefore, the instantaneous end-to-end SINR at PU$_i$ can be expressed using (25).
\begin{multline} 
\scalebox{0.95}{$\gamma_{i}^{AF} = \frac{\frac{\xi^2 \alpha P_{s} (1-\rho) P_{p_j}}{N_p^{'} (D_{i} D_j)^m} {\sum_{w=1}^{N_p}\sum_{w^{'}=1}^{N_p^{'}} \mid {h}_{i,w} \mid^2} {\mid {h}_{j,w^{'}} \mid}^2}{\frac{(1-\alpha)P_{s}\sum_{w=1}^{N_p}{\mid {h}_{i,w} \mid}^2 }{(D_{i})^m} +\frac{\xi^2 \alpha P_{s} \sum_{w=1}^{N_p}{\mid {h}_{i,w} \mid}^2 \sigma^2}{(D_{i})^m} + \sigma^2}=\frac{{C_i} {\sum_{w=1}^{N_p}\sum_{w^{'}=1}^{N_p^{'}} \mid {h}_{i,w} \mid^2} {\mid {h}_{j,w^{'}} \mid}^2}{{H_i} \sum_{w=1}^{N_p}{\mid {h}_{i,w} \mid}^4+{E_i} {\sum_{w=1}^{N_p}\sum_{w^{'}=1}^{N_p^{'}} \mid {h}_{i,w} \mid^2} {\mid {h}_{j,w^{'}} \mid}^2+F_i \sum_{w=1}^{N_p}{\mid {h}_{i,w} \mid}^2+1}$}
\end{multline}

where  C$_1$ = $\frac{\eta \rho P_{p_2} \alpha}{N_b \sigma^2 (D_{1})^m (D_{2})^m}$, C$_2$ = $\frac{\eta \rho P_{p_1} \alpha}{N_a \sigma^2 (D_{1})^m (D_{2})^m}$,    H$_1$ = $\frac{(1-\alpha) \eta \rho P_{p_1}}{N_a \sigma^2 (D_{1})^{2m}}$, H$_2$ = $\frac{(1-\alpha) \eta \rho P_{p_2}}{N_b \sigma^2 (D_{2})^{2m}}$, E$_1$ = $\frac{(1-\alpha) \eta \rho P_{p_2}}{N_b \sigma^2 (D_{1} D_{2})^m}$, E$_2$ = $\frac{(1-\alpha) \eta \rho P_{p_1}}{N_a \sigma^2 (D_{1} D_{2})^m}$, F$_1$ = $\frac{\rho \alpha \eta}{(1-\rho)(D_{1})^m}$, F$_2$ = $\frac{\rho \alpha \eta}{(1-\rho)(D_{2})^m}$.

Similarly, SU$_2$ is able to detect PU signal by considering SU signal as an interference [22]. The instantaneous SINR can be expressed applying (26).

  \begin{multline}
\scalebox{0.95}{$\gamma_{s,PU}^{AF} = \frac{{\frac{\alpha \eta \rho}{(D_3)^m \sigma^2  }}\mid{h}_{3}\mid ^2 \bigg(\sum_{n=1}^{N_a}\frac{P_{p_1}}{ (D_1)^m N_a} \mid{h}_{1,n}\mid ^2 + \sum_{p=1}^{N_b}\frac{P_{p_2}}{(D_2)^m N_b} \mid{h}_{2,p}\mid ^2 \bigg)}{{\frac{(1-\alpha)}{(D_3)^m \sigma^2 }}\eta \rho \mid{h}_{3}\mid ^2 \bigg(\sum_{n=1}^{N_a} \frac{P_{p_1}}{ (D_1)^m N_a } \mid{h}_{1,n}\mid ^2 +\sum_{p=1}^{N_b} \frac{P_{p_2}}{(D_2)^m N_b} \mid{h}_{2,p}\mid ^2 \bigg)+ \frac{\alpha \eta \rho \mid{h}_{3}\mid ^2}{(1-\rho) (D_3)^m}+1}$}\\\scalebox{0.95}{$=\frac{\mid{h}_{3}\mid ^2 \big(\sum_{n=1}^{N_a}{U_{3}} \mid{h}_{1,n}\mid ^2 +\sum_{p=1}^{N_b} {V_{3}} \mid{h}_{2,p}\mid ^2 \big)}{\mid{h}_{3}\mid ^2 \big(\sum_{n=1}^{N_a}{U_{1}} \mid{h}_{1,n}\mid ^2 +\sum_{p=1}^{N_b} {V_{1}} \mid{h}_{2,p}\mid ^2 \big)+U_2 \mid{h}_{3}\mid ^2  +1}$}
\end{multline}

  where U$_1$=${\frac{(1-\alpha)}{(D_3)^m \sigma^2}}\eta \rho \frac{P_{p_1}}{N_a (D_1)^m}$, V$_1$=${ \frac{(1-\alpha)}{(D_3)^m \sigma^2}}\eta \rho \frac{P_{p_2}}{N_b (D_2)^m}$, U$_2$=${\frac{\alpha}{(D_3)^m (1-\rho)}}\eta \rho$, U$_3$=$\frac{\alpha \eta \rho P_{p_1}}{(D_3 D_1)^m \sigma^2 N_a }$, V$_3$=$\frac{\alpha \eta \rho P_{p_2}}{(D_3 D_2)^m \sigma^2 N_b }$.
 
The PU signals being a strong one, SU$_2$ first decodes PU signals considering SU$_1$ signal as interference and  removes it from the received signal. Then it decodes its own signal in the presence of noise. The instantaneous SNR at SU$_2$ can be expressed in (27).
 \begin{align}
\scalebox{0.95}{$\gamma_{s}^{AF} = \frac{{\frac{(1-\alpha)}{(D_3)^m \sigma^2}}\eta \rho \mid{h}_{3}\mid ^2 \bigg(\sum_{n=1}^{N_a}\frac{P_{p_1}}{ (D_1)^m N_a } \mid{h}_{1,n}\mid ^2 +\sum_{p=1}^{N_b} \frac{P_{p_2}}{(D_2)^m N_b} \mid{h}_{2,p}\mid ^2 \bigg)}{ \frac{\alpha \eta \rho \mid{h}_{3}\mid ^2}{(1-\rho) (D_3)^m}+1}=\frac{\mid{h}_{3}\mid ^2 \big(\sum_{n=1}^{N_a}{U_{1}} \mid{h}_{1,n}\mid ^2 +\sum_{p=1}^{N_b} {V_{1}} \mid{h}_{2,p}\mid ^2 \big)}{U_2 \mid{h}_{3}\mid ^2  +1}$}
\end{align}

  Achievable rate at $PU_i$ is given by (based on the SINR)
\begin{align} 
\scalebox{0.95}{$R_{PU_i}^{(2,AF)}= \frac{1}{2} \log_2 \lbrace 1+ \gamma_{i}^{AF} \rbrace$}
\end{align}

The achievable rate for PU information decoding at $SU_2$ is as 
\begin{align} 
\scalebox{01}{$R_{s,PU}^{(2,AF)}= \frac{1}{2} \log_2 \lbrace 1+ \gamma_{s,PU}^{AF} \rbrace$}
\end{align}

The achievable rate at $SU_2$ is given by (based on the SNR at $SU_2$)
\begin{align} 
\scalebox{01}{$R_{SU_2}^{(2,AF)}= \frac{1}{2} \log_2 \lbrace 1+ \gamma_{s}^{AF} \rbrace$}
\end{align}

 \subsection{\textbf {Outage Probability Analysis}}

  \subsubsection{Outage Probability of Primary System}
  PU outage probability using AF relaying is determined as follows:
  \begin{equation}
\scalebox{1}{$\mathscr{P}_{out}^{(PU,AF)} = 1 - \overbrace{\bigg[\mathscr{P}\big\lbrace R_{PU_1}^{(2,AF)}\geq R_{PU}\big\rbrace \times \mathscr{P}\big\lbrace R_{PU_2}^{(2,AF)}\geq R_{PU}\big\rbrace\bigg]}^{\textbf{Success Probability}}$}
\end{equation}

Following (25), the probability of success due to data transmission between the link SU$_1$ to PU$_1$ is expressed as [21, Sec. 3.471.9, 8.352.1] 
\begin{multline} 
\scalebox{01}{$\mathscr{P}\{R_{PU_1}^{(2,AF)}\geq R_{PU}\}= \mathscr{P}\bigg[\frac{{C_1} \sum_{n=1}^{N_a}{\mid {h}_{1,n} \mid}^2 \sum_{p=1}^{N_b}{\mid {h}_{2,p} \mid}^2}{{H_1} \sum_{n=1}^{N_a}{\mid {h}_{1,n} \mid}^4+{E_1} \sum_{n=1}^{N_a}{\mid {h}_{1,n} \mid}^2 \sum_{p=1}^{N_b}{\mid {h}_{2,p} \mid}^2+F_1 \sum_{n=1}^{N_a}{\mid {h}_{1,n} \mid}^2+1}\geq u_1\bigg]$}\\\scalebox{0.95}{$= \mathscr{P}\bigg[\frac{{C_1} X Y}{{H_1} {X}^2+{E_1} X Y +F_1 X+1}\geq u_1\bigg]=1-\mathscr{P}\bigg[Y < \frac{F_1 u_1}{(C_1-E_1 u_1)} + \frac{H_1 u_1 X}{(C_1-E_1 u_1)} +\frac{u_1}{(C_1-E_1 u_1)X}\bigg]=\sum_{{q_a}=0}^{N_b-1}(N_b u_1)^{q_a}    $}\\ \scalebox{0.9}{$2  {\Bigg( {\frac{N_b u_1/(C_1-E_1 u_1)}{N_a+\frac{N_b H_1 u_1}{(C_1-E_1 u_1)}}}\Bigg)}^{(N_a+q-2l)/2} \frac{\exp\big(-\frac{N_b F_1{u_{1}}}{(C_1-E_1 u_1)}\big)\sum_{q=0}^{{q_a}}{{{q_a}}\choose{q}} \sum_{l=0}^{q}{{q}\choose{l}} {F_1}^{{q_a}-q} {H_1}^{q-l}}{{q_a}!  \Gamma(N_a) \big(1/N_a\big)^{N_a} {(C_1-E_1 u_1)}^{q_a}} K_{N_a+q-2l}\bigg(2\sqrt{\frac{N_b u_1}{(C_1-E_1 u_1)}\big({N_a+\frac{N_b H_1 u_1}{C_1-E_1 u_1}}}\big)\bigg)$}
\end{multline}

\textbf{Proof} : See Appendix D.

where $X$=$\sum_{n=1}^{N_a}{\mid {h}_{1,n} \mid}^2$, $Y$=$\sum_{p=1}^{N_b}{\mid {h}_{2,p} \mid}^2$. The symbol K$_v (.)$ indicates the modified Bessel function. Similarly, it is also possible to determine the probability of successful data transmission between the link SU$_1$ to PU$_2$.

Closed form expression of PU outage probability using AF relaying is given in Appendix H.

\subsubsection{Outage Probability Analysis of Secondary System}
SU outage probability, using AF relaying mechanism, can be determined as follows:

\begin{align} 
\scalebox{01}{$\mathscr{P}_{out}^{(SU,AF)} = 1 - \bigg[\underbrace{\mathscr{P}\big\lbrace R_{SU_2}^{(2,AF)}\geq R_{SU}|R_{s,PU}^{(2,AF)}\geq R_{PU}\big\rbrace}_{\textbf{Success of SU$_1$ $\rightarrow$ SU$_2$}} \times \underbrace{\mathscr{P}\big\lbrace R_{s,PU}^{(2,AF)}\geq R_{PU}\big\rbrace}_{\textbf{Successful decoding of PU signals at SU$_2$}}\bigg]$}
\end{align}
  
  Following (27), the probability of successful data transmission between the link SU$_1$ to SU$_2$ is expressed using (34) [21, Sec. 3.471.9, 8.352.2].
\begin{multline} 
\scalebox{0.95}{$\mathscr{P}( R_{SU_2}^{(2,AF)}\geq R_{SU})=\mathscr{P}\Bigg\{ \frac{\mid{h}_{3}\mid ^2 \Bigg({U_{1}}\sum_{n=1}^{N_a} \mid{h}_{1,n}\mid ^2 + {V_{1}} \sum_{p=1}^{N_b}\mid{h}_{2,p}\mid ^2 \Bigg)}{U_2 \mid{h}_{3}\mid ^2  +1} \geq u_4\Bigg\}$}\\\scalebox{0.95}{$=\sum_{p_a=0}^{N_b-1} \sum_{r=0}^{p_a}{{p_a}\choose{r}} \sum_{l=0}^{r}(-1)^{l}{{r}\choose{l}}\Big(\frac{u_4 U_2 N_b}{V_1}\Big)^{p_a -r} \frac{\bigg(\frac{u_4 N_b}{ V_{1}}\bigg)^{r-l}      \bigg(\frac{U_{1}  N_b}{V_{1}}\bigg)^{l} \Gamma(N_a+l)}{p_a! \big\{({1/N_a})^{N_a}\big\} \bigg( {N_a } -\frac{U_{1} N_b}{V_{1}}\bigg)^{N_a+l}  \Gamma(N_a)}\exp\Big(-\frac{u_4 U_2 N_b}{V_1}\Big) $}\\\scalebox{0.95}{$ \Bigg[\frac{2{\big(\frac{u_4 N_b}{V_1 m_k}\big)}^{\frac{m_k+l-r}{2}}K_{m_k+l-r}\bigg(2\sqrt{\frac{u_4 {N_b}m_k}{V_1}}\bigg)}{\{({1/m_k})^{m_k}\} \Gamma(m_k)}- \sum_{j=0}^{N_a+l-1}\exp\bigg\{-\bigg(\frac{u_4 U_2}{U_1}\bigg)\bigg( {N_a } -\frac{U_{1} N_b}{V_{1}}\bigg)\bigg\} $}\\\scalebox{0.95}{$ \frac{\bigg( {N_a } -\frac{U_{1} N_b}{V_{1}}\bigg)^j \sum_{t_l=0}^{j}{{j}\choose{t_l}}\bigg(\frac{u_4 U_2}{U_1}\bigg)^{j-t_l} \bigg(\frac{u_4}{ U_1}\bigg)^{t_l}}       {j! \{({1/m_k})^{m_k}\} \Gamma(m_k)}   2{\bigg(\frac{u_4 N_a}{U_1 m_k}\bigg)}^{\frac{m_k+l-r-t_l}{2}}K_{m_k+l-r-t_l}\bigg(2\sqrt{\frac{u_4 {N_a}m_k}{U_1}}\bigg)\Bigg]  $}\\\scalebox{0.95}{$ +\sum_{p_a=0}^{N_a-1}\frac{\bigg({\frac{u_{4} {N_{a}}}{U_{1}}}\bigg)^{p_a}  ({m_k})^{ m_k} \exp\Big(-\frac{u_4 U_2 N_a}{U_1}\Big)}{p_a! \Gamma(m_k) } \sum_{r=0}^{p_a}{{p_a}\choose{r}} {(U_2)}^{p_a-r} 2  {\Bigg(      {\frac{u_{4}N_a}{U_{1}m_k}}\Bigg)}^{(m_k -r)/2}K_{m_k - r}\bigg(2\sqrt{{{\frac{u_{4} {N_{a}}m_k}{U_{1} }}}}\bigg)$}
\end{multline}

\textbf{Proof} : See Appendix E.


Similar to (34), the probability of success for the decoding of the PU information in presence of SU interference and noise can be written as
\small{\begin{multline} 
\scalebox{0.95}{$\mathscr{P}\big\lbrace R_{s,PU}^{(2,AF)}\geq R_{PU}\big\rbrace=\mathscr{P}\bigg\lbrace\frac{Z \big({U_{3}} X_1 + {V_{3}} Y_1 \big)}{Z \big({U_{1}}X_1 + {V_{1}} Y_1 \big)+U_2 Z +1} \geq u_1 \bigg\rbrace= \begin{cases}
\scalebox{0.95}{$\mathscr{P}\big\lbrace Z \big({S_{1}} X_1 + {S_{2}} Y_1 \big)\geq u_s U_2 z+u_s \rbrace,\hspace*{0.25cm} {\text for,} \hspace*{0.15cm} u_s < \frac{\alpha}{(1-\alpha)}$}\\
\scalebox{0.9}{$ 0,$} \hspace*{5cm} otherwise .\\
\end{cases}$}
\\=\begin{cases}\scalebox{0.95}{$\sum_{p_a=0}^{N_b-1} \sum_{r=0}^{p_a}{{p_a}\choose{r}} \sum_{l=0}^{r}(-1)^{l}{{r}\choose{l}}\Big(\frac{u_s U_2 N_b}{S_2}\Big)^{p_a -r}\exp\Big(-\frac{u_s U_2 N_b}{S_2}\Big)  \frac{\bigg(\frac{u_s N_b}{ S_{2}}\bigg)^{r-l}      \bigg(\frac{S_{1}  N_b}{S_{2}}\bigg)^{l} \Gamma(N_a+l)}{p_a! \big\{({1/N_a})^{N_a}\big\} \bigg( {N_a } -\frac{S_{1} N_b}{S_{2}}\bigg)^{N_a+l}  \Gamma(N_a)}$}\\\scalebox{0.95}{$\Bigg[\frac{2{\big(\frac{u_s N_b}{S_2 m_k}\big)}^{\frac{m_k+l-r}{2}}}{\{({1/m_k})^{m_k}\} \Gamma(m_k)}K_{m_k+l-r}\bigg(2\sqrt{\frac{u_s {N_b}m_k}{S_2}}\bigg)  - \sum_{j=0}^{N_a+l-1}\exp\bigg\{-\bigg(\frac{u_s U_2}{S_1}\bigg)\bigg( {N_a } -\frac{S_{1} N_b}{S_{2}}\bigg)\bigg\}  $}\\\scalebox{0.95}{$ 2{\Big(\frac{u_s N_a}{S_1 m_k}\Big)}^{\frac{m_k+l-r-t_l}{2}} \frac{\bigg( {N_a } -\frac{S_{1} N_b}{S_{2}}\bigg)^j \sum_{t_l=0}^{j}{{j}\choose{t_l}}\bigg(\frac{u_s U_2}{S_1}\bigg)^{j-t_l} \bigg(\frac{u_s}{ S_1}\bigg)^{t_l}}       {j! \{({1/m_k})^{m_k}\} \Gamma(m_k)} 
 K_{m_k+l-r-t_l}\bigg(2\sqrt{\frac{u_s {N_a}m_k}{S_1}}\bigg)\Bigg] $}\\\scalebox{0.95}{$+\sum_{p_a=0}^{N_a-1}\frac{\bigg({\frac{u_{s} {N_{a}}}{S_{1}}}\bigg)^{p_a}  ({m_k})^{m_k} \exp\Big(-\frac{u_s U_2 N_a}{S_1}\Big)}{p_a! \Gamma(m_k) } \sum_{r=0}^{p_a}{{p_a}\choose{r}} {(U_2)}^{p_a-r} 2  {\Bigg(      {\frac{u_{s} N_a}{S_{1} m_k}}\Bigg)}^{(m_k -r)/2} K_{m_k - r}\bigg(2\sqrt{{{\frac{u_{s} {N_{a}}m_k}{S_{1} }}}}\bigg),$}\\\hspace*{10.15cm} {\text for,} \hspace*{0.15cm} u_s < \frac{\alpha}{(1-\alpha)}\\
\scalebox{0.86}{$ 0,$} \hspace*{10.2cm} otherwise .
\end{cases}\\\vspace*{-3cm}
\end{multline}}

where $S_1= {\frac{1}{(D_3)^m \sigma^2}}\eta \rho \frac{P_{p_1}}{N_a (D_1)^m}$, $S_2={ \frac{1}{(D_3)^m \sigma^2}}\eta \rho \frac{P_{p_2}}{N_a (D_2)^m}$, $u_s=\frac{u_1}{\alpha-u_1 (1-\alpha)}$. 
\begin{table}[b]
\caption{Set of numerical values of necessary parameters}
  \centering
    \begin{tabular}{*{29}{c}}
    \hline
    Name & Value\\
    \hline
    \hline
    PU target rate of information transmission ($R_{PU}$)  & 0.2 bps/Hz \\
    \hline
    SU target rate of information transmission ($R_{SU}$)  & 1 bps/Hz \\
    \hline
    Distance between PU$_1$ and PU$_2$ (L)  & 20 m \\ 
    \hline 
    Distance between SU$_1$ and SU$_2$ (D$_3$) & 10 m \\
    \hline
    PU transmit power ($P_{p_1}$=$P_{p_2}$=$P_{p}$) & -23 dBm  (Fig. 3 to Fig. 5), -50 dBm to -15 dBm (Fig. 6)\\
    \hline
    Average noise power ($n_{pu}$=$n_{su}$=$n_p$) & -100 dBm [11]\\
    \hline
     Path loss exponent ($ m $) & 2.7 \\  
    \hline
     Energy conversion efficiency ($ \eta $) & 0.9 \\  
    \hline
    Power splitting factor ($\rho$)   &  0.9\\
    \hline
    Power sharing factor ($\alpha$)   &  0.81\\
    \hline
    $D_1=D_4$ \& $D_2=D_5$  &  $L/2$\\
    \hline
     \end{tabular}
  \end{table}

Finally, the closed form SU outage expression using AF relaying can be determined using (34) and (35).

Closed form expression of SU outage probability using AF relaying is given in Appendix M.


\section{Numerical Results and Discussions}
Numerical values for the system parameters used are enlisted in Table II. The constraint of the maximum PU outage limit (say 0.1) is found to be met at $\alpha$=0.81 by both AF and DF relaying for N$_a$=2. This value of $\alpha$ is used for the results shown in Fig. 4, Fig. 5 and Fig. 6. Considering the practicability of the number of transmitting antennas on mobile devices we set N$_b$=1 in our simulation results.

\begin{figure}
  \centering
  \includegraphics[width=0.8\linewidth]{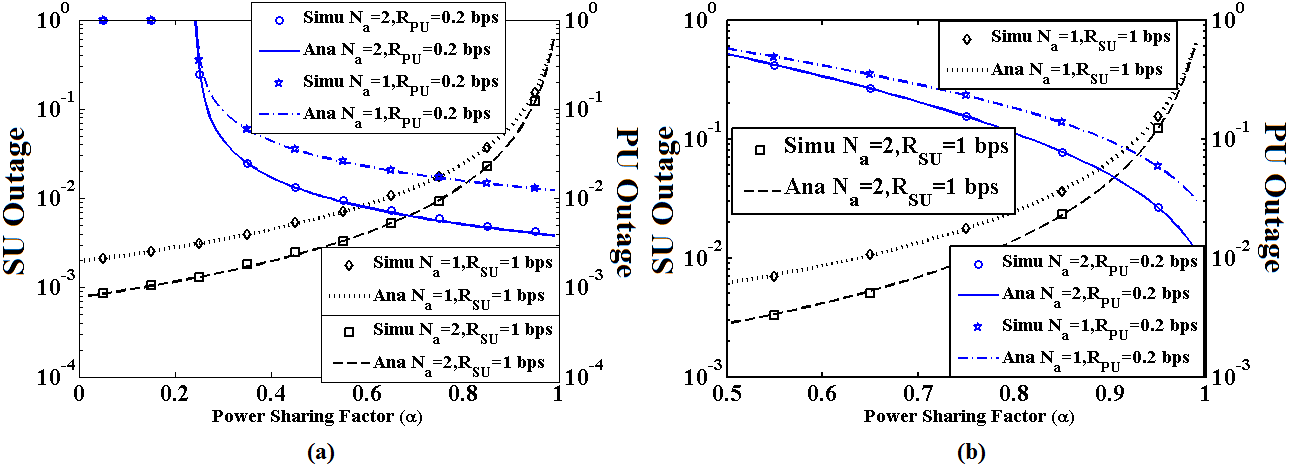}
  \caption{{Outage Probability vs. $\alpha$: (a) DF relaying (b) AF relaying}}
\end{figure}

The outage performances of both PU and SU systems are shown as variation in power-sharing factor ($\alpha$) for both DF and AF relaying in Fig. 3.a and Fig. 3.b. It is clearly seen that the analytical results match perfectly with the simulation results. For the small value of $\alpha$, the higher fraction of the harvesting power is allocated to SU information transmission and small fraction of power is used for DF relaying of PU information. Thus SU outage is found to be low while PU outage to be high. The reverse situation is to observed at high value of $\alpha$. The small value of $\alpha$ (close to 0) causes PU outage of AF relaying same to that of DF relaying and SU performance is found to be the worst by following (33)-(35). Since 50$\%$ or more power is assigned to PU information transmission to detect PU signals perfectly in presence of SU interference, therefore the minimum value of $\alpha$ is shown as 0.5 for AF relaying in Fig. 3.b. The threshold limit of 10$\%$ PU outage is achieved at $\alpha$=0.29 and N$_a$=1 for DF relaying and the same outage limit is achieved at $\alpha$=0.89 and N$_a$=1 for AF relaying. If the number of antennas is increased then PU performance is improved accordingly. If N$_a$ is increased from 1 to 2, about 46$\%$ improvement (reduction) in PU outage is observed for DF relaying and about 39$\%$ improvement of PU outage is observed for AF relaying. If N$_a$ is increased from 1 to 2, about 58$\%$ improvement of SU outage is observed at $\alpha$=0.27 and R$_{SU}$=1 bps for DF relaying and about 42$\%$ improvement of SU outage is observed at $\alpha$=0.81 and R$_{SU}$=1 bps for AF relaying. 


\begin{figure}
\centering
  \centering                       
  \includegraphics[width=0.8\textwidth]{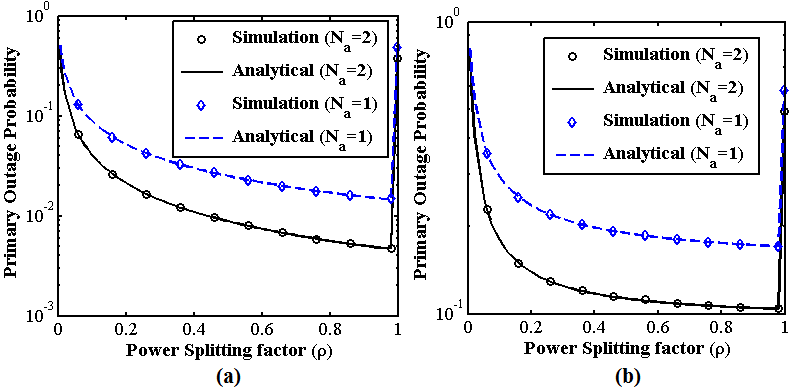}
  \caption{{PU Outage Probability vs. $\rho$}: (a) DF relaying, (b) AF relaying}
 \end{figure}%
\begin{figure}
  \centering
 \includegraphics[width=0.8\textwidth]{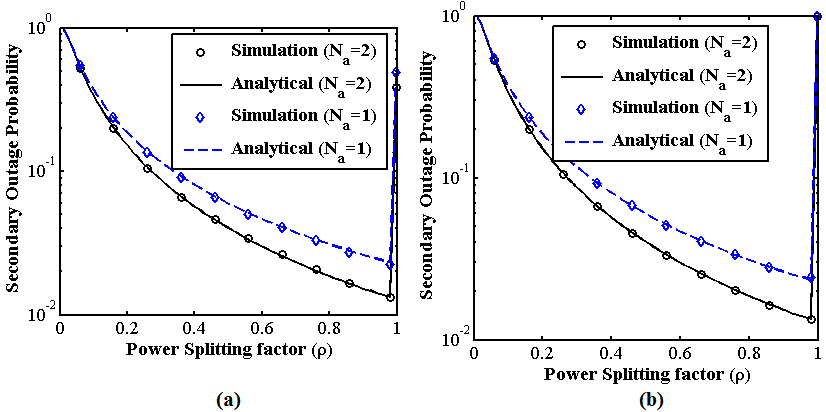}
  \caption{{SU Outage Probability vs. $\rho$}: (a) DF relaying, (b) AF relaying}
\end{figure}
   
   Fig. 4. and 5. show the outage performances of PU and SU, respectively with respect to power splitting factor ($\rho$) for both DF and AF relaying mechanisms. As depicted in the figures, the outage performances of both PU and SU are very poor for $\rho \rightarrow 0$ and $\rho \rightarrow 1$. Initially, the performances of PU and SU outage are improved with the increase in the value of $\rho$, and they attain their minimum values of outage at the optimal value of $\rho^{*}$. Thereafter, when the value of $\rho$ is increased further, it leads to an increase in both PU and SU outage. The reason behind the characteristics of this graphical plot can be explained as follows. Initially, when $\rho$ value is very low, the energy harvested at SU$_{1}$ is insufficient to broadcast the information at the BC phase and effectively the outage performance of both PU and SU are very poor. When the value of $\rho$ increases, the harvesting energy at SU$_1$ is also increased and consequently both PU and SU outage performances improve accordingly. Further degradation on the outage performances are found due to major power allocation for EH and less power allocation for decoding the information received from PU to SU signal transmission. Performance of PU outage is significantly improved for DF relaying as compared to AF relaying mechanism whereas the SU outage performances are almost same for both the relaying mechanisms with respect to $\rho$. The overall outage performances for both PU and SU are improved with the increase in the number of antennas used at PU nodes.
      \begin{figure}
  \centering
  \includegraphics[width=0.8\linewidth]{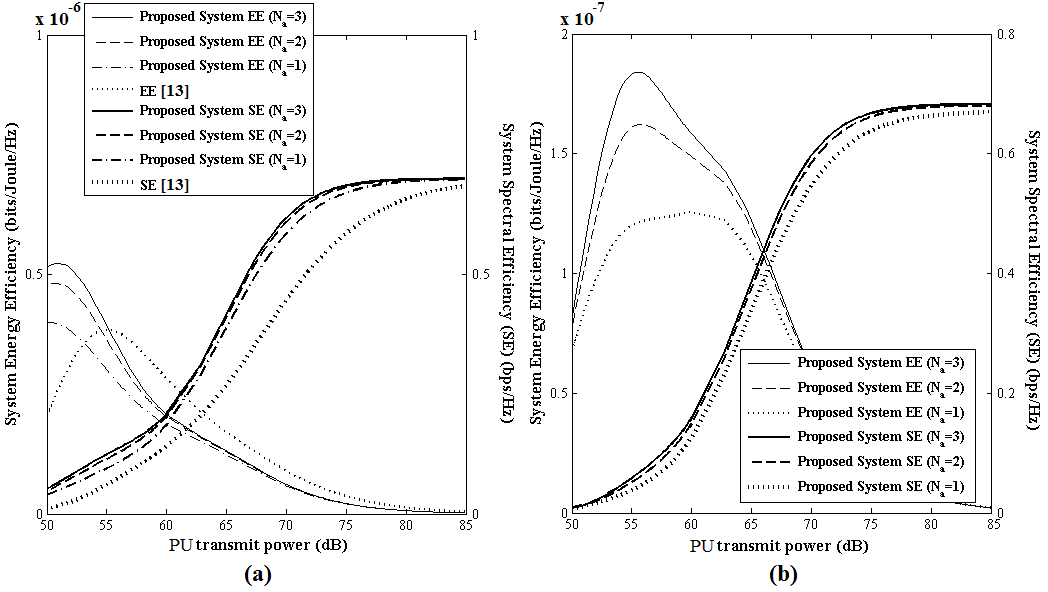}
  \caption{{System SE and EE vs. total PU transmit power : (a) DF relaying (b) AF relaying}}
\end{figure}

Now SE and EE of the system can be defined as [13]
\begin{align}
\eta_{SE}=\scalebox{01}{$2 \times (1-\mathscr{P}_{out}^{PU})\times{R_{PU}}\times\frac{T}{2T}+ (1-\mathscr{P}_{out}^{SU})\times{R_{SU}}\times\frac{T}{2T}$}       
\end{align}

EE can be defined as the ratio of SE to the total energy consumed for PUs transmission [23].

The SE and EE performances of the DF and AF relaying protocols are shown with respect to the total PU transmit power\footnote{The transmit power is normalized with respect to average channel noise power.} in Fig. 6.a and Fig. 6.b, respectively. The performance of the proposed system using AF and DF relaying schemes are compared with similar system supporting two-way PU and one-way SU communication simultaneously [13]. As shown in figure, SE is improved and EE is deteriorated with the increase in PU transmit power. The characteristics of this graphical plot may be explained as follows. As transmission power of PUs are increased, the PU and SU outage probabilities are decreased accordingly following (14)-(18), (21), (23), (32), (34) and (35) for DF and AF relaying mechanisms, respectively. As inverse relationship is maintained between SE and the outage probability of both PU and SU system, the SE is improved with the increase in the transmission power of PU. Less increment of SE as compared to more energy consumption leads to a degradation of EE with the increasing value of PUs transmit power. Our proposed system using DF relaying mechanism is more efficient as compared to AF relaying mechanism and similar model of two-way PU and one-way SU communication. In terms of SE, our proposed system using DF relaying mechanism is performing 224$\%$, 196$\%$ and 142$\%$ better as compared to [13] for N$_a$=3, N$_a$=2 and N$_a$=1, respectively at 52 dB transmit power. Our system using AF relaying is performing 34$\%$, 31$\%$ and 22$\%$ better in terms of SE as compared to [13] for given value of N$_a$=3, N$_a$=2 and N$_a$=1, respectively at 70 dB transmit power. In terms of EE, our proposed system using DF relaying mechanism is performing 62$\%$, 48$\%$ and 21$\%$ better as compared to [13] and on the other hand, the performance of [13] is 56$\%$, 60$\%$ and 69$\%$ better as compared to the proposed system using AF relaying for given value of N$_a$=3, N$_a$=2 and N$_a$=1, respectively at 52 dB transmit power. However, the proposed model is less energy efficient at high transmit power as compared to [13].

\section{Conclusions}
This paper has investigated the scope of SWIPT enabled IoT communication, on the licensed spectrum using overlay spectrum sharing mode of cognitive radio networks. Transmitting IoD node harvests energy from the information bearing RF signals of both eNB and UE transmission and this energy is used in relaying between eNB and UE. Apart from the close match between the analytical and simulation results on outage experiences in IoD and cellular systems, it is also to be noted that DF relaying mechanism is more efficient over AF relaying in terms of EE and AF relaying is more sensitive to the impact of the number of antennas used by the cellular system. The proposed network architecture and the spectrum sharing model can be extended further (i) to analyse the outage for more realistic non-linear RF-EH model, (ii) to improve the secrecy of the relay assisted cellular communication using the PUs' signals for friendly jamming in addition to EH and (iii) to study game theoretic modelling of the possible negotiations between cellular users and multiple IoT device pairs for efficient trading of the available radio resources of the former.

\section{Important Integrals and their solutions}
\subsubsection*{VII.1} [21, Sec. 3.471.9]
\begin{equation}
\int\limits_{0}^{\infty}{x}^{v-1} \text{exp}\bigg(-\frac{\beta}{x}-{\gamma}x\bigg) dx= 2 \Bigg(\frac{\beta}{\gamma}\Bigg)^{\frac{v}{2}} K_v(2\sqrt{\beta \gamma}), \big[\textbf{Re}(\beta) > 0,\textbf{Re}(\gamma) > 0\big]
\end{equation}

\subsubsection*{VII.2} [21, Sec. 3.351.3]
\begin{equation}
\int\limits_{0}^{\infty}{x}^{N_a} \text{exp}(-{\mu}x) dx= N_a!{\mu}^{-N_a-1}, \big[\textbf{Re}(\mu) > 0\big]
\end{equation}

\subsubsection*{VII.3} [21, Sec. 3.381.1]
\begin{equation}
\int\limits_{0}^u\frac{{x}^{v-1} \text{exp}(-{\mu}x)}{\Gamma(\mu)} dx= \frac{{\mu}^{-v} \gamma(v, \mu u)}{\Gamma(\mu)}, \big[\textbf{Re}(v) > 0\big]
\end{equation}

\subsubsection*{VII.4} [21, Sec. 3.381.3]
\begin{equation}
\int\limits_{u}^{\infty}\frac{{x}^{v-1} \text{exp}(-{\mu}x)}{\Gamma(\mu)} dx= \frac{{\mu}^{-v} \Gamma(v, \mu u)}{\Gamma(\mu)}, \big[u > 0,\textbf{Re}(\mu) > 0\big]
\end{equation}

\subsubsection*{VII.5} [21, Sec. 8.352.1]
\begin{equation}
\gamma(1+N_a, x) = N_a! \Bigg[1- \exp(-x) \Bigg(\sum_{r=0}^{N_a} \frac{{x}^r}{r!} \Bigg)\Bigg], {\bigg[N_a=0,1,...\bigg]}
\end{equation}

\subsubsection*{VII.6} [21, Sec. 8.352.2]
\begin{equation}
\Gamma(1+N_a, x) = N_a!  \exp(-x) \Bigg(\sum_{r=0}^{N_a} \frac{{x}^r}{r!} \Bigg),  {\bigg[N_a=0,1,...\bigg]}
\end{equation}

\subsubsection*{VII.7} [Taylor Series Expansion]
\begin{equation}
\exp(-x) = \sum_{n=0}^{\infty} (-1)^n \frac{x^n}{n!}
\end{equation}

\subsubsection*{VII.8} [Taylor Series Expansion]
\begin{equation}
\exp(x) = \sum_{n=0}^{\infty} \frac{x^n}{n!}
\end{equation}

\subsubsection*{VII.9} [Binomial Series Expansion]
\begin{equation}
(x+a)^n = \sum_{r=0}^{n} {{n}\choose{r}} x^{n-r} a^r
\end{equation}

\subsubsection{Appendix A: Proof of (14)} [21, Sec. 3.381.1, 3.381.3, 8.350.1, 8.352.1].
   \begin{multline} 
\scalebox{0.8}{$\mathscr{P}\lbrace Q_1 \rbrace=\int\limits_{\frac{u_{1}}{A_1}}^\infty \frac{{x_1}^{N_a-1} \text{exp}\big(-\dfrac{x_1}{1/N_a}\big)}{\Gamma(N_a) \big(1/N_a\big)^{N_a}} \Bigg\{\int\limits_{\frac{u_{2}}{A_2}}^\infty \frac{{y_1}^{N_b-1} \text{exp}\big(-\dfrac{y_1}{1/N_b}\big)}{\Gamma(N_b) \big(1/N_b\big)^{N_b}}dy_{1}\Bigg\}dx_{1}-\int\limits_{\frac{u_{1}}{A_1}}^{\frac{u_{3}-u_{1}}{A_1}} \frac{{x_1}^{N_a-1} \text{exp}\big(-\dfrac{x_1}{1/N_a}\big)}{\Gamma(N_a) \big(1/N_a\big)^{N_a}} \Bigg\{\int\limits_{\frac{u_{2}}{A_2}}^{\frac{u_{3}-A_1 x_{1}}{A_2}} \frac{{y_1}^{N_b-1} \text{exp}\big(-\dfrac{y_1}{1/N_b}\big)}{\Gamma(N_b) \big(1/N_b\big)^{N_b}}dy_{1}\Bigg\}dx_{1}$}\\\scalebox{0.84}{$=\int\limits_{\frac{u_{1}}{A_1}}^\infty \frac{{x_1}^{N_a-1} \text{exp}\big(-\dfrac{x_1}{1/N_a}\big)}{\Gamma(N_a) \big(1/N_a\big)^{N_a}} \Bigg\{\int\limits_{\frac{u_{2}}{A_2}}^\infty \frac{{y_1}^{N_b-1} \text{exp}\big(-\dfrac{y_1}{1/N_b}\big)}{\Gamma(N_b) \big(1/N_b\big)^{N_b}}dy_{1}\Bigg\}dx_{1}-\int\limits_{\frac{u_{1}}{A_1}}^{\frac{u_{3}-u_{1}}{A_1}} \frac{{x_1}^{N_a-1} \text{exp}\big(-\dfrac{x_1}{1/N_a}\big)}{\Gamma(N_a) \big(1/N_a\big)^{N_a}} \Bigg\{\int\limits_{0}^{\frac{u_{3}-A_1 x_{1}}{A_2}} \frac{{y_1}^{N_b-1} \text{exp}\big(-\dfrac{y_1}{1/N_b}\big)}{\Gamma(N_b) \big(1/N_b\big)^{N_b}}dy_{1}\Bigg\}dx_{1}$}\\\scalebox{0.85}{$+\int\limits_{\frac{u_{1}}{A_1}}^{\frac{u_{3}-u_{1}}{A_1}} \frac{{x_1}^{N_a-1} \text{exp}\big(-\dfrac{x_1}{1/N_a}\big)}{\Gamma(N_a) \big(1/N_a\big)^{N_a}} \Bigg\{\int\limits_{0}^{\frac{u_{2}}{A_2}} \frac{{y_1}^{N_b-1} \text{exp}\big(-\dfrac{y_1}{1/N_b}\big)}{\Gamma(N_b) \big(1/N_b\big)^{N_b}}dy_{1}\Bigg\}dx_{1}$}\\\scalebox{0.73}{$=\dfrac{1}{\Gamma(N_a)\Gamma(N_b)}{\Gamma\Bigg(N_a,{{N_a u_1}/{A_1}}\Bigg)\Gamma\Bigg(N_b,{{N_b u_2}/{A_2}}\Bigg)}-\int\limits_{\frac{u_{1}}{A_1}}^{\frac{u_{3}-u_{1}}{A_1}} \frac{{x_1}^{N_a-1} \text{exp}\big(-\dfrac{x_1}{1/N_a}\big)}{\Gamma(N_a) \Gamma(N_b) \big(1/N_a\big)^{N_a}} \gamma\Bigg(N_b,N_b \dfrac{u_{3}-A_1 x_{1}}{A_2}\Bigg) dx_{1}+\int\limits_{\frac{u_{1}}{A_1}}^{\frac{u_{3}-u_{1}}{A_1}} \frac{{x_1}^{N_a-1} \text{exp}\big(-\dfrac{x_1}{1/N_a}\big)}{\Gamma(N_a) \Gamma(N_b) \big(1/N_a\big)^{N_a}} \gamma\Bigg(N_b,N_b \dfrac{u_{2}}{A_2}\Bigg) dx_{1}
   $}\\\scalebox{0.75}{$=\dfrac{1}{\Gamma(N_a)\Gamma(N_b)}{\Gamma\Bigg(N_a,{{N_a u_1}/{A_1}}\Bigg)\Gamma\Bigg(N_b,{{N_b u_2}/{A_2}}\Bigg)}$}\\\scalebox{0.73}{$-\dfrac{1}{\Gamma(N_a) \big(1/N_a\big)^{N_a}}\int\limits_{\frac{u_{1}}{A_1}}^{\frac{u_{3}-u_{1}}{A_1}} {{x_1}^{N_a-1} \text{exp}\big(-\dfrac{x_1}{1/N_a}\big)} \Bigg[1-\frac{\exp\bigg(-N_b \dfrac{u_{3}-A_1 x_{1}}{A_2}\bigg)}{{q_a}!} \sum_{q_a=0}^{N_b-1} {N_b}^{q_a} \big(\dfrac{u_{3}-A_1 x_{1}}{A_2}\big)^{{q_a}}\Bigg] dx_{1}+\int\limits_{\frac{u_{1}}{A_1}}^{\frac{u_{3}-u_{1}}{A_1}} \frac{{x_1}^{N_a-1} \text{exp}\big(-\dfrac{x_1}{1/N_a}\big)}{\Gamma(N_a) \Gamma(N_b) \big(1/N_a\big)^{N_a}} \gamma\Bigg(N_b,N_b \dfrac{u_{2}}{A_2}\Bigg) dx_{1}$}\\\scalebox{0.8}{$ \text{(Using Binomial Series Expansion, we can write as follows)}$}\\\scalebox{0.77}{$=\dfrac{1}{\Gamma(N_a)\Gamma(N_b)}{\Gamma\Bigg(N_a,{{N_a u_1}/{A_1}}\Bigg)\Gamma\Bigg(N_b,{{N_b u_2}/{A_2}}\Bigg)}-\dfrac{1}{\Gamma(N_a) \big(1/N_a\big)^{N_a}}\Bigg\{\int\limits_{0}^{\frac{u_{3}-u_{1}}{A_1}} {{x_1}^{N_a-1} \text{exp}\big(-\dfrac{x_1}{1/N_a}\big)}dx_1-\int\limits_{0}^{\frac{u_{1}}{A_1}} {{x_1}^{N_a-1} \text{exp}\big(-\dfrac{x_1}{1/N_a}\big)}dx_1\Bigg\}$}\\\scalebox{0.77}{$+\sum_{q_a=0}^{N_b-1}\frac{\exp\big(-\frac{N_b {u_{3}}}{A_2}\big)\sum_{q=0}^{{q_a}}(-1)^{q}{{q_a}\choose{q}} {u_3}^{{q_a}-q} {A_1}^q}{{q_a}!  \Gamma(N_a) \big(1/N_a\big)^{N_a} \big(1/N_b\big)^{{q_a}} {A_2}^{q_a}}\Bigg\{\int\limits_{0}^{\frac{u_{3}-u_{1}}{A_1}} {{x_1}^{N_a +q-1} \text{exp}\big\{-\big(N_a-\frac{A_1 N_b}{A_2}\big) x_1\big\}}dx_1-\int\limits_{0}^{\frac{u_{1}}{A_1}} {{x_1}^{N_a +q-1} \text{exp}\big\{-\big(N_a-\frac{A_1 N_b}{A_2}\big) x_1\big\}}dx_1\Bigg\} $}\\\scalebox{0.8}{$+\dfrac{\gamma\Big(N_b,N_b \dfrac{u_{2}}{A_2}\Big)}{\Gamma(N_a)\Gamma(N_b) \big(1/N_a\big)^{N_a}}\Bigg\{\int\limits_{0}^{\frac{u_{3}-u_{1}}{A_1}} {{x_1}^{N_a-1} \text{exp}\big(-\dfrac{x_1}{1/N_a}\big)}dx_1-\int\limits_{0}^{\frac{u_{1}}{A_1}} {{x_1}^{N_a-1} \text{exp}\big(-\dfrac{x_1}{1/N_a}\big)}dx_1\Bigg\}$}\\ \scalebox{0.85}{$=\dfrac{1}{\Gamma(N_a)\Gamma(N_b)}{\Gamma\Bigg(N_a,{{N_a u_1}/{A_1}}\Bigg)\Gamma\Bigg(N_b,{{N_b u_2}/{A_2}}\Bigg)}-\Bigg\{\dfrac{1}{\Gamma({N_a})}-\dfrac{\gamma\Big(N_b,N_b \dfrac{u_{2}}{A_2}\Big)}{\Gamma(N_a) \Gamma(N_b)}\Bigg\}\Big\{\gamma\big(N_a,N_a\dfrac{u_{3}-u_{1}}{A_1}\big)-\gamma\big(N_a,N_a\dfrac{u_1}{A_1}\big)\Big\}$}\\\scalebox{0.85}{$+\sum_{q_a=0}^{N_b-1}\dfrac{\exp\big(-\frac{N_b {u_{3}}}{A_2}\big)\sum_{q=0}^{{q_a}}(-1)^{q}{{{q_a}}\choose{q}} {u_3}^{{q_a}-q} {A_1}^q}{{q_a}!  \Gamma(N_a) \big(1/N_a\big)^{N_a} \big(1/N_b\big)^{{q_a}} \Big(N_a-\frac{A_1 N_b}{A_2}\Big)^{(N_a+q)}{A_2}^{q_a}}\Big\{\gamma\big[N_a+q, \big(N_a-\frac{A_1 N_b}{A_2}\big)\dfrac{u_{3}-u_{1}}{A_1}\big]-\gamma\big[N_a+q, \big(N_a-\frac{A_1 N_b}{A_2}\big)\dfrac{u_{1}}{A_1}\big]\Big\} $}
\end{multline}

\subsubsection{Appendix B: Proof of (17)} [21, Sec. 8.350.1, 8.352.1].
\begin{multline} 
\scalebox{0.88}{$\mathscr{P}\Bigg\{X_1 \leq \dfrac{k^{'}}{(a_1 X_1+b_1 Y_1)} \Bigg\}=\mathscr{P}\Bigg( Y_1\leq \dfrac{k^{'}}{b_{1} X_{1}}-\dfrac{a_{1} X_{1}}{b_{1}}\Bigg)=\int\limits_0^{\sqrt{\dfrac{k^{'}}{a_1}}} \frac{{x_1}^{N_a-1} \text{exp}\big(-\dfrac{x_1}{1/N_a}\big)}{\Gamma(N_a) \big(1/N_a\big)^{N_a}} \Bigg\{\int\limits_{0}^{\frac{k^{'}}{b_{1} x_{1}}-\dfrac{a_{1} x_{1}}{b_{1}}} \frac{{y_1}^{N_b-1} \text{exp}\big(-\dfrac{y_1}{1/N_b}\big)}{\Gamma(N_b) \big(1/N_b\big)^{N_b}}dy_{1}\Bigg\}dx_{1}$}\\\scalebox{0.88}{$= \dfrac{1}{\Gamma(N_b)}\int\limits_0^{\sqrt{\frac{k^{'}}{a_{1}}}} \frac{\{{x_{1}}^{( N_a-1)}\} \text{exp}\big(-\frac{x_{1}} {{1/N_a}}\big)}{\{({1/N_a})^{N_a}\} \Gamma(N_a)} {\gamma\Bigg(N_b,{{N_b\Big[\dfrac{k^{'} }{b_{1} x_{1}}-\dfrac{a_{1} x_{1}}{b_{1}}\Big]}}\Bigg)}dx_{11}$}\\\scalebox{0.88}{$= \dfrac{1}{\Gamma(N_b)}\int\limits_0^{\sqrt{\frac{k^{'}}{a_{1}}}} \frac{\{{x_{1}}^{(N_a-1)}\} \text{exp}\big(-\frac{x_{1}} {{1/N_a}}\big)}{\{({1/N_a})^{N_a}\} \Gamma(N_a)} 
\Gamma( N_b) \Bigg[1- \exp\Big\{-{{N_b\Big(\dfrac{k^{'}}{b_{1} x_{1}}-\dfrac{a_{1} x_{1}}{b_{1}}\Big)}}\Big\}\sum_{p_a=0}^{N_b-1}\dfrac{\bigg\{{N_b {\Big(\dfrac{k^{'}}{b_{1} x_{1}}-\dfrac{a_{1} x_{1}}{b_{1}}\Big)}}\bigg\}^{p_a}}{p_a!}\Bigg]dx_{1}$}\\ \scalebox{0.8}{$\text{Using Binomial Series Expansion, we can write as follows},$}\\\scalebox{0.74}{$ =\dfrac{1}{\Gamma(N_a)}{\gamma\Bigg(N_a,\dfrac{\sqrt{\dfrac{k^{'} }{a_{1}}}}{1/N_a}\Bigg)}-\Bigg[\sum_{p_a=0}^{N_b-1}\dfrac{\big({N_b}\big)^{p_a}}{p_a! ({1/N_a})^{N_a} \Gamma(N_a)}  \sum_{r=0}^{p_a}(-1)^{r}{{p_a}\choose{r}} \bigg(\dfrac{k^{'}}{b_{1}}\bigg)^{p_a-r} \bigg(\dfrac{a_{1}}{b_{1}}\bigg)^{r} \int\limits_0^{\sqrt{\frac{k^{'} }{a_{1}}}}  (x_{1})^{2r+N_a -p_a-1 }  \text{exp}\Bigg\{-{\dfrac{\Big(\dfrac{k^{'} N_b}{b_{1}}\Big)}{x_{1}}}-\Bigg({N_a-\dfrac{a_{1}N_b}{b_{1}}} \Bigg)x_{1}\Bigg\} dx_{1}\Bigg]
$}\\ \scalebox{0.8}{$\text{Using Taylor Series Expansion, we can write as follows},$}\\ \scalebox{0.74}{$ =\dfrac{1}{\Gamma(N_a)}{\gamma\Bigg(N_a,\dfrac{\sqrt{\dfrac{k^{'} }{a_{1}}}}{1/N_a}\Bigg)}-\Bigg[\sum_{p_a=0}^{N_b-1}\dfrac{\big({N_b}\big)^{p_a}}{p_a! ({1/N_a})^{N_a} \Gamma(N_a)}  \sum_{r=0}^{p_a}(-1)^{r}{{p_a}\choose{r}} \bigg(\dfrac{k^{'}}{b_{1}}\bigg)^{p_a-r} \bigg(\dfrac{a_{1}}{b_{1}}\bigg)^{r}  \sum_{l=0}^{\infty} \dfrac{(-1)^l}{l!}  \int\limits_0^{\sqrt{\frac{k^{'}}{a_{1}}}}  (x_{1})^{2r+N_a -p_a-1 } \Bigg\{{\dfrac{\Big(\dfrac{k^{'} N_b}{b_{1}}\Big)}{x_{1}}}+\Bigg({N_a-\dfrac{a_{1}N_b}{b_{1}}} \Bigg)x_{1}\Bigg\}^l  dx_{1}\Bigg]$}\\ \scalebox{0.8}{$\text{Using Binomial Series Expansion, we can write as follows},$}\\ \scalebox{0.76}{$ =\dfrac{1}{\Gamma(N_a)}{\gamma\Bigg(N_a,\dfrac{\sqrt{\dfrac{k^{'} }{a_{1}}}}{1/N_a}\Bigg)}-\Bigg[\sum_{p_a=0}^{N_b-1}\dfrac{\big({N_b}\big)^{p_a}}{p_a! ({1/N_a})^{N_a} \Gamma(N_a)}  \sum_{r=0}^{p_a}(-1)^{r}{{p_a}\choose{r}} \bigg(\dfrac{k^{'}}{b_{1}}\bigg)^{p_a-r} \bigg(\dfrac{a_{1}}{b_{1}}\bigg)^{r}  \sum_{l=0}^{\infty} \dfrac{(-1)^l}{l!}\sum_{t_l=0}^{l}{{l}\choose{t_l}} {\Big(\dfrac{k^{'} N_b}{b_{1}}\Big)}^{l-t_l} \Bigg({N_a-\dfrac{a_{1}N_b}{b_{1}}} \Bigg)^{t_l} $}\\ \scalebox{0.84}{$\int\limits_0^{\sqrt{\frac{k^{'}}{a_{1}}}}  (x_{1})^{2r+ N_a -p_a-l+2t_l-1}   dx_{1}\Bigg]$}\\ \scalebox{0.84}{$ =\dfrac{1}{\Gamma(N_a)}{\gamma\Bigg(N_a,\dfrac{\sqrt{\dfrac{k^{'} }{a_{1}}}}{1/N_a}\Bigg)}-\Bigg[\sum_{p_a=0}^{N_b-1}\dfrac{\big({N_a}\big)^{N_a}\big({N_b}\big)^{p_a}}{p_a!  \Gamma(N_a)}\sum_{r=0}^{p_a}(-1)^{r}{{p_a}\choose{r}} \bigg(\dfrac{k^{'}}{b_{1}}\bigg)^{p_a-r} \bigg(\dfrac{a_{1}}{b_{1}}\bigg)^{r}  \sum_{l=0}^{\infty} \dfrac{(-1)^l}{l! (2r+ N_a -p_a-l+2t_l)}$}\\ \scalebox{0.84}{$ \sum_{t_l=0}^{l}{{l}\choose{t_l}} {\Big(\dfrac{k^{'} N_b}{b_{1}}\Big)}^{l-t_l} \Bigg({N_a-\dfrac{a_{1}N_b}{b_{1}}} \Bigg)^{t_l}  \Bigg(\sqrt{\dfrac{k^{'}}{a_{1}}}\Bigg)^{2r+ N_a -p_a-l+2t_l}   \Bigg]$}
\end{multline}

\subsubsection{Appendix C: Proof of (23)}[21, Sec. 3.471.9, 8.352.2], [24].
\begin{multline}
\scalebox{0.83}{$\mathscr{P}( R_{SU_2}^{(2,DF)}\geq R_{SU})=\mathscr{P}\Big( Z\geq \dfrac{u_4}{c(a_1 X_1 +b_1 Y_1)}\Big)=\mathscr{P}\bigg(Y_{1}\geq \dfrac{u_4}{c b_{1} Z}-\dfrac{a_{1} X_{1}}{b_{1}}\bigg)
$}\\\scalebox{0.71}{$ =\int\limits_0^{\infty} \frac{\{{z}^{(m_k-1)}\} \text{exp}(-\frac{z} {{1/m_k}})}{\{({1/m_k})^{m_k}\} \Gamma(m_k)}\bigg[ \int\limits_0^{\frac{u_{4}}{c a_{1} z}}\frac{\{{x_{1}}^{(N_a-1)}\} \text{exp}(-\frac{x_{1}} {{1/N_a}})}{\{({1/N_a})^{N_a}\} \Gamma(N_a)} \bigg\{\int\limits_{\frac{u_4}{c b_{1} z}-\dfrac{a_{1} x_{1}}{b_{1}}}^{\infty}\frac{\{{y_{1}}^{( N_b-1)}\} \text{exp}(-\frac{y_{1}} {{1/N_b}})}{\{({1/N_b})^{N_b}\} \Gamma(N_b)} dy_{1}\bigg\}dx_{1}\bigg]  dz
+\int\limits_0^{\infty} \frac{\{{z}^{(m_k-1)}\} \text{exp}(-\frac{z} {{1/m_k}})}{\{({1/m_k})^{m_k}\} \Gamma(m_k)}\bigg[ \int\limits_{\frac{u_{4}}{c a_{1} z}}^{\infty} \frac{\{{x_{1}}^{(N_a-1)}\} \text{exp}(-\frac{x_{1}} {{1/N_a}})}{\{({1/N_a})^{N_a}\} \Gamma(N_a)} dx_{1}\bigg]  dz$}\\\scalebox{0.87}{$=\int\limits_0^{\infty} \frac{\{{z}^{(m_k-1)}\} \text{exp}(-\frac{z} {{1/m_k}})}{\{({1/m_k})^{m_k}\} \Gamma(m_k)}\bigg[ \int\limits_0^{\frac{u_{4}}{c a_{1} z}}\frac{\{{x_{1}}^{(N_a-1)}\} \text{exp}(-\frac{x_{1}} {{1/N_a}})}{\{({1/N_a})^{N_a}\} \Gamma(N_a)} \dfrac{\Gamma\Big(N_b,\dfrac{u_4 {N_b}}{c b_{1} z }-\dfrac{a_{1} x_{1} N_b}{b_{1}}\Big)}{\Gamma(N_b)}dx_{1}\bigg]  dz + I$}\\\scalebox{0.83}{$
=\int\limits_0^{\infty} \frac{\{{z}^{(m_k-1)}\} \text{exp}(-\frac{z} {{1/m_k}})}{\{({1/m_k})^{m_k}\} \Gamma(m_k)}\bigg[ \int\limits_0^{\frac{u_{4}}{c a_{1} z}}\frac{\{{x_{1}}^{(N_a-1)}\} \text{exp}(-\frac{x_{1}} {{1/N_a}})}{\{({1/N_a})^{N_a}\} \Gamma(N_a)} \bigg\{\exp\Big(-\dfrac{u_4 {N_b}}{c b_{1} z }+\dfrac{a_{1} x_{1} N_b}{b_{1}}\Big)\sum_{p_a=0}^{N_b-1}\dfrac{\bigg(\dfrac{u_4 {N_b}}{c b_{1} z }-\dfrac{a_{1} x_{1} N_b}{b_{1}}\bigg)^{p_a}}{p_a!}\bigg\}dx_{1}\bigg]  dz + I
$}\\\scalebox{0.78}{$ =\sum_{p_a=0}^{N_b-1}  \dfrac{\Big(\dfrac{N_b}{b_1}\Big)^{p_a}}{p_a!  ({1/N_a})^{ N_a} \Gamma(N_a)} \sum_{r=0}^{p_a}(-1)^{r}{{p_a}\choose{r}}  \bigg(\dfrac{u_4}{c}\bigg)^{p_a-r} \big({a_{1}}\big)^{r}\int\limits_0^{\infty} \frac{\{{z}^{(m_k-p_a+r-1)}\} \text{exp}(-\frac{z} {{1/m_k}}-\frac{u_4 {N_b}}{c b_{1} z })}{\{({1/m_k})^{ m_k}\} \Gamma(m_k)}   \int\limits_0^{\frac{u_{4}}{c a_{1} z}} {x_{1}}^{N_a+r-1} \exp\bigg\{-{\Big(N_a-\dfrac{a_{1} N_b}{b_{1}}\Big) x_{1}}\bigg\} dx_{1}dz  + I $}\\{\text{(Following Binomial Series Expansion)}}\\\scalebox{0.77}{$ =\sum_{p_a=0}^{N_b-1}  \dfrac{\Big(\dfrac{N_b}{b_1}\Big)^{p_a}}{p_a!  ({1/N_a})^{ N_a} \Gamma(N_a)} \sum_{r=0}^{p_a}(-1)^{r}{{p_a}\choose{r}}  \bigg(\dfrac{u_4}{c}\bigg)^{p_a-r} \big({a_{1}}\big)^{r}\int\limits_0^{\infty} \frac{\{{z}^{(m_k-p_a+r-1)}\} \text{exp}\bigg(-\dfrac{z} {{1/m_k}}-\dfrac{u_4 {N_b}}{c b_{1} z }\bigg)}{\{({1/m_k})^{m_k}\} \Gamma(m_k)}\dfrac{1}{{\bigg(N_a-\frac{a_{1} N_b}{b_{1}}\bigg)}^{N_a+r}}\gamma\Bigg(N_a+r,\frac{u_4  \bigg({N_a}-\dfrac{a_{1}{N_b}}{b_{1}}\bigg)}{cz a_{1}}\Bigg) dz  + I$}\\\scalebox{0.82}{$ =\sum_{p_a=0}^{N_b-1}  \dfrac{\Big(\dfrac{N_b}{b_1}\Big)^{p_a}}{p_a!  ({1/N_a})^{N_a} \Gamma(N_a)} \sum_{r=0}^{p_a}(-1)^{r}{{p_a}\choose{r}}  \bigg(\dfrac{u_4}{c}\bigg)^{p_a-r} \big({a_{1}}\big)^{r}\int\limits_0^{\infty} \frac{\{{z}^{(m_k-p_a+r-1)}\} \text{exp}\bigg(-\dfrac{z} {{1/m_k}}-\dfrac{u_4 {N_b}}{c b_{1} z }\bigg)}{\{({1/m_k})^{m_k}\} \Gamma(m_k)}\dfrac{\Gamma(N_a+r)}{{\bigg(N_a-\frac{a_{1}N_b}{b_{1}}\bigg)}^{N_a+r}}$}\\\scalebox{0.78}{$  \Bigg\{1- \dfrac{1}{j!} \exp\bigg(-\frac{u_4  \bigg({N_a}-\dfrac{a_{1}{N_b}}{b_{1}}\bigg)}{cz a_{1}}\bigg) \sum_{j=0}^{N_a+r-1}\bigg[\frac{u_4  \bigg({N_a}-\dfrac{a_{1}{N_b}}{b_{1}}\bigg)}{cz a_{1}}\bigg]^j \Bigg\}  dz  + I$}\\\scalebox{0.78}{$ =  \sum_{p_a=0}^{N_b-1}  \dfrac{\Big(\dfrac{N_b}{b_1}\Big)^{p_a}}{p_a!  ({1/N_a})^{N_a} \Gamma(N_a)} \sum_{r=0}^{p_a}(-1)^{r}{{p_a}\choose{r}}  \bigg(\dfrac{u_4}{c}\bigg)^{p_a-r} \big({a_{1}}\big)^{r}\frac{1}{{\bigg(N_a-\frac{a_{1}N_b}{b_{1}}\bigg)}^{N_a+r}}\dfrac{\Gamma{(N_a +r)}}{\{({1/m_k})^{m_k}\} \Gamma(m_k)}\bigg[2\bigg\{\frac{u_4 N_b}{c b_{1} m_k} \bigg\}^{(m_k-p_a+r)/2} $}\\\scalebox{0.78}{$ K_{m_k-p_a+r}\bigg\{2\sqrt{\frac{u_4 {N_b} m_k}{c b_{1}}}\bigg\}   - \sum_{j=0}^{N_a+r-1} \dfrac{1}{j!} \bigg[\frac{u_4  \bigg({N_a}-\dfrac{a_{1}{N_b}}{b_{1}}\bigg)}{c a_{1}} \bigg]^{j} \int\limits_0^{\infty} {\{{z}^{(m_k-p_a+r-j-1)}\}}\exp\bigg\{-\dfrac{z} {{1/m_k}}-\dfrac{u_4 {N_b}}{c b_{1} z }-\frac{u_4  \bigg({N_a}-\dfrac{a_{1}{N_b}}{b_{1}}\bigg)}{cz a_{1}}\bigg\}dz \bigg]+ I$}\\\scalebox{0.76}{$ =  \sum_{p_a=0}^{N_b-1}  \dfrac{\Big(\dfrac{N_b}{b_1}\Big)^{p_a}}{p_a!  ({1/N_a})^{N_a} \Gamma(N_a)} \sum_{r=0}^{p_a}(-1)^{r}{{p_a}\choose{r}}  \bigg(\dfrac{u_4}{c}\bigg)^{p_a-r} \big({a_{1}}\big)^{r}\frac{1}{{\bigg(N_a-\frac{a_{1}N_b}{b_{1}}\bigg)}^{N_a+r}}\dfrac{\Gamma{(N_a +r)}}{\{({1/m_k})^{m_k}\} \Gamma(m_k)}\bigg[2\bigg\{\frac{u_4 N_b}{c b_{1} m_k} \bigg\}^{(m_k-p_a+r)/2} $}\\\scalebox{0.76}{$ K_{m_k-p_a+r}\bigg\{2\sqrt{\frac{u_4 {N_b} m_k}{c b_{1}}}\bigg\}   - \sum_{j=0}^{N_a+r-1} \dfrac{1}{j!} \bigg[\frac{u_4  \bigg({N_a}-\dfrac{a_{1}{N_b}}{b_{1}}\bigg)}{c a_{1}} \bigg]^{j}  2\bigg\{\frac{u_4  \bigg(N_a-\frac{a_{1}N_b}{b_{1}}\bigg)}{c a_{1} m_k}+\frac{u_4 N_b}{c b_{1} m_k}\bigg\}^{\frac{(m_k-p_a+r-j)}{2}}$}\\\scalebox{0.76}{$ K_{m_k-p_a+r-j}\bigg\{2\sqrt{\frac{u_4  m_k \bigg({N_a}-\frac{a_{1}{N_b}}{b_{1}}\bigg)}{c a_{1}}+\frac{u_4 {N_b} m_k}{c b_{1} }}\bigg\}  \Bigg]+\sum_{p_a=0}^{N_a-1}\dfrac{\bigg({\dfrac{u_{4} {N_{a}}}{c a_{1}}}\bigg)^{p_a}  ({m_k})^{m_k}}{p_a! \Gamma(m_k)} 2  {\Bigg(      {\dfrac{u_{4} N_a}{c a_{1} m_k}}\Bigg)}^{(m_k - p_a)/2}K_{m_k - p_a}\bigg(2\sqrt{{{\dfrac{u_{4} {N_{a}} m_k}{c a_{1} }}}}\bigg)$}
\end{multline}

I is possible to determine using (49) [21, Sec. 3.471.9, 8.352.2].

\begin{multline}
\scalebox{0.84}{$I=\int\limits_0^{\infty} \frac{\{{z}^{(m_k-1)}\} \text{exp}(-\frac{z} {{1/m_k}})}{\{({1/m_k})^{m_k}\} \Gamma(m_k)}\bigg[ \int\limits_{\frac{u_{4}}{c a_{1} z}}^{\infty} \frac{\{{x_{1}}^{(N_a-1)}\} \text{exp}(-\frac{x_{1}} {{1/N_a}})}{\{({1/N_a})^{N_a}\} \Gamma(N_a)} dx_{1}\bigg]  dz =\dfrac{1}{\Gamma{(N_a)}}   \int\limits_0^{\infty} \frac{\{{z}^{(m_k-1)}\} \text{exp}(-\frac{z} {{1/m_k}})}{\{({1/m_k})^{m_k}\} \Gamma(m_k)}{\Gamma\Bigg(N_a,{{\dfrac{u_{4}N_a}{c a_{1} z}}}\Bigg)}dz$}\\\scalebox{0.78}{$ =\int\limits_0^{\infty} \frac{\{{z}^{(m_k-1)}\} \text{exp}(-\frac{z} {{1/m_k}})}{\{({1/m_k})^{m_k}\} \Gamma(m_k)}\Bigg[\exp\Big(-{\dfrac{u_{4} {N_{a}}}{c a_{1}z}}\Big)\sum_{p_a=0}^{N_a-1}\dfrac{\bigg({\dfrac{u_{4} {N_{a}}}{c a_{1}z}}\bigg)^{p_a}}{p_a!}\Bigg]dz=\sum_{p_a=0}^{N_a-1}\dfrac{\bigg({\dfrac{u_{4} {N_{a}}}{c a_{1}}}\bigg)^{p_a}  ({m_k})^{m_k}}{p_a! \Gamma(m_k) }\int\limits_0^{\infty} {\big(z\big)}^{m_k - p_a -1} \exp\Bigg(-\dfrac{z} {{1/m_k}}-\dfrac{1}{ z}{\dfrac{u_{4} {N_{a}}}{c a_{1}}}\Bigg)dz$}\\\scalebox{0.87}{$=\sum_{p_a=0}^{N_a-1}\dfrac{\bigg({\dfrac{u_{4} {N_{a}}}{c a_{1}}}\bigg)^{p_a}  ({m_k})^{m_k}}{p_a! \Gamma(m_k)} 2  {\Bigg(      {\dfrac{u_{4} N_a}{c a_{1} m_k}}\Bigg)}^{(m_k - p_a)/2}K_{m_k - p_a}\bigg(2\sqrt{{{\dfrac{u_{4} {N_{a}} m_k}{c a_{1} }}}}\bigg)$}
\end{multline}

\subsubsection{Appendix D: Proof of (32)}[21, Sec. 3.471.9, 8.352.1].
\begin{multline} 
\scalebox{0.83}{$\mathscr{P}\bigg[Y < \dfrac{F u_1}{(C-E u_1)} + \dfrac{H u_1 X}{(C-E u_1)} +\dfrac{u_1}{(C-E u_1)X}\bigg]=\int\limits_{0}^\infty \frac{{x}^{N_a-1} \text{exp}\big(-\dfrac{x}{1/N_a}\big)}{\Gamma(N_a) \big(1/N_a\big)^{N_a}} \Bigg\{\int\limits_{0}^{\bigg[\frac{F u_1}{(C-E u_1)} + \frac{H u_1 x}{(C-E u_1)} +\frac{u_1}{(C-E u_1)x}\bigg]} \frac{{y}^{N_b-1} \text{exp}\big(-\dfrac{y}{1/N_b}\big)}{\Gamma(N_b) \big(1/N_b\big)^{N_b}}dy\Bigg\}dx$}\\\scalebox{0.85}{$=\int\limits_{0}^\infty \frac{{x}^{N_a-1} \text{exp}\big(-\dfrac{x}{1/N_a}\big)}{\Gamma(N_a) \Gamma(N_b) \big(1/N_a\big)^{N_a}} \gamma \bigg\{N_b,\dfrac{F u_1 N_b}{(C-E u_1)} + \dfrac{H u_1  N_b x}{(C-E u_1)} +\dfrac{u_1  N_b}{(C-E u_1)x}\bigg\}dx$}\\\scalebox{0.7}{$=\dfrac{1}{\Gamma(N_a) \big(1/N_a\big)^{N_a}}\int\limits_{0}^{\infty} {{x}^{N_a-1} \text{exp}\big(-\dfrac{x}{1/N_a}\big)} \Bigg[1-\frac{\exp\bigg\{-N_b \bigg(\dfrac{F u_1}{(C-E u_1)} + \dfrac{H u_1 x}{(C-E u_1)} +\dfrac{u_1}{(C-E u_1)x}\bigg)\bigg\}}{{q_a}!} \sum_{q_a=0}^{N_b-1} {N_b}^{q_a} \bigg(\dfrac{F u_1}{(C-E u_1)} + \dfrac{H u_1 x}{(C-E u_1)} +\dfrac{u_1}{(C-E u_1)x}\bigg)^{{q_a}}\Bigg] dx$}\\\scalebox{0.8}{$ \text{(Using Binomial Series Expansion, we can write as follows)}$}\\\scalebox{0.86}{$=1-\sum_{q_a=0}^{N_b-1}(N_b u_1)^{q_a}\frac{\exp\big(-\frac{N_b F{u_{1}}}{(C-E u_1)}\big)\sum_{q=0}^{q_a}{{q_a}\choose{q}} \sum_{l=0}^{q}{{q}\choose{l}} F^{{q_a}-q} H^{q-l}}{{q_a}!  \Gamma(N_a) \big(1/N_a\big)^{N_a} {(C-E u_1)}^{q_a}}\Bigg\{\int\limits_{0}^{\infty} {{x}^{N_a +q-2l-1} \text{exp}\big\{-N_a x-\dfrac{N_b H u_1 x}{(C-E u_1)}-\dfrac{N_b u_1}{(C-E u_1)x}\big\}}dx\Bigg\} $}\\ \scalebox{0.79}{$=1-\sum_{q_a=0}^{N_b-1}(N_b u_1)^{q_a}\frac{\exp\big(-\frac{N_b F{u_{1}}}{(C-E u_1)}\big)\sum_{q=0}^{{q_a}}{{{q_a}}\choose{q}} \sum_{l=0}^{q}{{q}\choose{l}} F^{{q_a}-q} H^{q-l}}{{q_a}!  \Gamma(N_a) \big(1/N_a\big)^{N_a} {(C-E u_1)}^{q_a}} 2  {\Bigg( {\dfrac{N_b u_1/(C-E u_1)}{N_a+\frac{N_b H u_1}{(C-E u_1)}}}\Bigg)}^{(N_a+q-2l)/2}K_{N_a+q-2l}\bigg(2\sqrt{\dfrac{N_b u_1}{(C-E u_1)}\big({N_a+\frac{N_b H u_1}{C-E u_1}}}\big)\bigg) $}
\end{multline}

\subsubsection{Appendix E: Proof of (34)} [21, Sec. 3.471.9, 8.352.2], [24].
\begin{multline}
\scalebox{0.83}{$\mathscr{P}( R_{SU_2}^{(2,AF)}\geq R_{SU})=\mathscr{P}\Bigg\{ \dfrac{ Z \Bigg({U_{1}} X_1 + {V_{1}} Y_1 \Bigg)}{U_2 Z  +1} \geq u_4\Bigg\}=\mathscr{P}\bigg(Y_{1}\geq \dfrac{u_4}{Z V_{1}}+\dfrac{u_4 U_2}{V_1}-\dfrac{U_{1} X_{1}}{V_{1}}\bigg)
$}\\\scalebox{0.76}{$ =\int\limits_0^{\infty} \frac{\{{z}^{(m_k-1)}\} \text{exp}(-\frac{z} {{1/m_k}})}{\{({1/m_k})^{m_k}\} \Gamma(m_k)}\bigg[ \int\limits_0^{\frac{u_4}{z U_1}+\dfrac{u_4 U_2}{U_1}}\frac{\{{x_{1}}^{(N_a-1)}\} \text{exp}(-\frac{x_{1}} {{1/N_a}})}{\{({1/N_a})^{N_a}\} \Gamma(N_a)} \bigg\{\int\limits_{\frac{u_4}{z V_{1}}+\frac{u_4 U_2}{V_1}-\frac{U_{1} x_{1}}{V_{1}}}^{\infty}\frac{\{{y_{1}}^{( N_b-1)}\} \text{exp}(-\frac{y_{1}} {{1/N_b}})}{\{({1/N_b})^{N_b}\} \Gamma(N_b)} dy_{1}\bigg\}dx_{1}\bigg]  dz
$}\\\scalebox{0.76}{$+\int\limits_0^{\infty} \frac{\{{z}^{(m_k-1)}\} \text{exp}(-\frac{z} {{1/m_k}})}{\{({1/m_k})^{m_k}\} \Gamma(m_k)}\bigg[ \int\limits_{\frac{u_4}{z U_1}+\dfrac{u_4 U_2}{U_1}}^{\infty} \frac{\{{x_{1}}^{(N_a-1)}\} \text{exp}(-\frac{x_{1}} {{1/N_a}})}{\{({1/N_a})^{N_a}\} \Gamma(N_a)} dx_{1}\bigg]  dz$}\\\scalebox{0.86}{$=\int\limits_0^{\infty} \frac{\{{z}^{(m_k-1)}\} \text{exp}(-\frac{z} {{1/m_k}})}{\{({1/m_k})^{m_k}\} \Gamma(m_k)}\bigg[ \int\limits_0^{\frac{u_4}{z U_1}+\dfrac{u_4 U_2}{U_1}}\frac{\{{x_{1}}^{(N_a-1)}\} \text{exp}(-\frac{x_{1}} {{1/N_a}})}{\{({1/N_a})^{N_a}\} \Gamma(N_a)}  \dfrac{\Gamma\Big(N_b,\dfrac{u_4 N_b}{z V_{1}}+\dfrac{u_4 U_2 N_b}{V_1}-\dfrac{U_{1} x_{1} N_b}{V_{1}}\Big)}{\Gamma(N_b)}dx_{1}\bigg]  dz + I^{'}$}\\\scalebox{0.86}{$
=\int\limits_0^{\infty} \frac{\{{z}^{(m_k-1)}\} \text{exp}(-\frac{z} {{1/m_k}})}{\{({1/m_k})^{m_k}\} \Gamma(m_k)}\bigg[ \int\limits_0^{\frac{u_4}{z U_1}+\dfrac{u_4 U_2}{U_1}}\frac{\{{x_{1}}^{(N_a-1)}\} \text{exp}(-\frac{x_{1}} {{1/N_a}})}{\{({1/N_a})^{N_a}\} \Gamma(N_a)}  \bigg\{\exp\Big(-\dfrac{u_4 N_b}{z V_{1}}-\dfrac{u_4 U_2 N_b}{V_1}+\dfrac{U_{1} x_{1} N_b}{V_{1}}\Big)$}\\\scalebox{0.87}{$\sum_{p_a=0}^{N_b-1}\dfrac{\bigg(\dfrac{u_4 N_b}{z V_{1}}+\dfrac{u_4 U_2 N_b}{V_1}-\dfrac{U_{1} x_{1} N_b}{V_{1}}\bigg)^{p_a}}{p_a!}\bigg\}dx_{1}\bigg]  dz + I^{'}$}\\\scalebox{0.87}{$
=\exp\Big(-\dfrac{u_4 U_2 N_b}{V_1}\Big)\int\limits_0^{\infty} \dfrac{\{{z}^{(m_k-1)}\} \text{exp}\Big(-\dfrac{z} {{1/m_k}}-\dfrac{u_4 N_b}{z V_{1}}\Big)}{\{({1/m_k})^{m_k}\} \Gamma(m_k)}\bigg[ \int\limits_0^{\frac{u_4}{z U_1}+\frac{u_4 U_2}{U_1}}\frac{\{{x_{1}}^{(N_a-1)}\} \text{exp}\Big(-\dfrac{x_{1}} {{1/N_a}}+\dfrac{U_{1} x_{1} N_b}{V_{1}}\Big)}{\{({1/N_a})^{N_a}\} \Gamma(N_a)}  $}\\\scalebox{0.85}{$\bigg\{\sum_{p_a=0}^{N_b-1}\dfrac{\bigg(\dfrac{u_4 N_b}{z V_{1}}+\dfrac{u_4 U_2 N_b}{V_1}-\dfrac{U_{1} x_{1} N_b}{V_{1}}\bigg)^{p_a}}{p_a!}\bigg\}dx_{1}\bigg]  dz + I^{'}$}\\{\text{(Using Binomial Series Expansion, we can write as follows)}}\\\scalebox{0.8}{$
 =\sum_{p_a=0}^{N_b-1} \sum_{r=0}^{p_a}{{p_a}\choose{r}} \sum_{l=0}^{r}(-1)^{l}{{r}\choose{l}} \Big(\dfrac{u_4 U_2 N_b}{V_1}\Big)^{p_a -r}\exp\Big(-\dfrac{u_4 U_2 N_b}{V_1}\Big) \dfrac{\bigg(\dfrac{u_4 N_b}{ V_{1}}\bigg)^{r-l}      \bigg(\dfrac{U_{1}  N_b}{V_{1}}\bigg)^{l}}{p_a!}\int\limits_0^{\infty} \dfrac{\{{z}^{(m_k-r+l-1)}\} \text{exp}\Big(-\dfrac{z} {{1/m_k}}-\dfrac{u_4 N_b}{z V_{1}}\Big)}{\{({1/m_k})^{m_k}\} \Gamma(m_k)} $}\\\scalebox{0.83}{$ \bigg[ \int\limits_0^{\frac{u_4}{z U_1}+\frac{u_4 U_2}{U_1}}\frac{\{{x_{1}}^{(N_a+l-1)}\} \text{exp}\Big(-\dfrac{x_{1}} {{1/N_a}}+\dfrac{U_{1} x_{1} N_b}{V_{1}}\Big)}{\{({1/N_a})^{N_a}\} \Gamma(N_a)} dx_{1}\bigg]  dz + I^{'}$}\\\scalebox{0.8}{$ =\sum_{p_a=0}^{N_b-1} \sum_{r=0}^{p_a}{{p_a}\choose{r}} \sum_{l=0}^{r}(-1)^{l}{{r}\choose{l}} \Big(\dfrac{u_4 U_2 N_b}{V_1}\Big)^{p_a -r}\exp\Big(-\dfrac{u_4 U_2 N_b}{V_1}\Big) \dfrac{\bigg(\dfrac{u_4 N_b}{ V_{1}}\bigg)^{r-l}      \bigg(\dfrac{U_{1}  N_b}{V_{1}}\bigg)^{l}}{p_a!}\int\limits_0^{\infty} \dfrac{\{{z}^{(m_k-r+l-1)}\} \text{exp}\Big(-{z} {{m_k}}-\dfrac{u_4 N_b}{z V_{1}}\Big)}{\{({1/m_k})^{m_k}\} \Gamma(m_k)} $}\\\scalebox{0.83}{$  \frac{\gamma\Big\{N_a+l,\bigg(\dfrac{u_4}{z U_1}+\dfrac{u_4 U_2}{U_1}\bigg)\bigg( {N_a } -\dfrac{U_{1} N_b}{V_{1}}\bigg)\Big\}}{\big\{({1/N_a})^{N_a}\big\} \bigg( {N_a } -\dfrac{U_{1} N_b}{V_{1}}\bigg)^{N_a+l}  \Gamma(N_a)}   dz + I^{'}$}\\\scalebox{0.83}{$ =\sum_{p_a=0}^{N_b-1} \sum_{r=0}^{p_a}{{p_a}\choose{r}} \sum_{l=0}^{r}(-1)^{l}{{r}\choose{l}} \Big(\dfrac{u_4 U_2 N_b}{V_1}\Big)^{p_a -r}\exp\Big(-\dfrac{u_4 U_2 N_b}{V_1}\Big) \dfrac{\bigg(\dfrac{u_4 N_b}{ V_{1}}\bigg)^{r-l}      \bigg(\dfrac{U_{1}  N_b}{V_{1}}\bigg)^{l} \Gamma(N_a+l)}{p_a!  \big\{({1/N_a})^{N_a}\big\} \bigg( {N_a } -\dfrac{U_{1} N_b}{V_{1}}\bigg)^{N_a+l}  \Gamma(N_a)} $}\\\scalebox{0.8}{$  \Bigg[1- \exp\bigg\{-\bigg(\dfrac{u_4}{z U_1}+\dfrac{u_4 U_2}{U_1}\bigg)\bigg( {N_a } -\dfrac{U_{1} N_b}{V_{1}}\bigg)\bigg\} \sum_{j=0}^{N_a+l-1} \dfrac{\bigg[\bigg(\dfrac{u_4}{z U_1}+\dfrac{u_4 U_2}{U_1}\bigg)\bigg( {N_a } -\dfrac{U_{1} N_b}{V_{1}}\bigg)\bigg]^j}{j!} \Bigg]$}\\\scalebox{0.8}{$ \int\limits_0^{\infty} \dfrac{\{{z}^{(m_k-r+l-1)}\} \text{exp}\Big(-{m_k z} -\dfrac{u_4 N_b}{z V_{1}}\Big)}{\{({1/m_k})^{m_k}\} \Gamma(m_k)}  dz  + I^{'}$}\\\scalebox{0.8}{$ =  \sum_{p_a=0}^{N_b-1} \sum_{r=0}^{p_a}{{p_a}\choose{r}} \sum_{l=0}^{r}(-1)^{l}{{r}\choose{l}} \Big(\dfrac{u_4 U_2 N_b}{V_1}\Big)^{p_a -r}\exp\Big(-\dfrac{u_4 U_2 N_b}{V_1}\Big) \dfrac{\bigg(\dfrac{u_4 N_b}{ V_{1}}\bigg)^{r-l}      \bigg(\dfrac{U_{1}  N_b}{V_{1}}\bigg)^{l} \Gamma(N_a+l)}{p_a!  \big\{({1/N_a})^{N_a}\big\} \bigg( {N_a } -\dfrac{U_{1} N_b}{V_{1}}\bigg)^{N_a+l}  \Gamma(N_a)} $}\\\scalebox{0.8}{$  \Bigg[\dfrac{2{\big(\dfrac{u_4 N_b}{V_1 m_k}\big)}^{\dfrac{m_k+l-r}{2}}K_{m_k+l-r}\bigg(2\sqrt{\dfrac{u_4 {N_b}m_k}{V_1}}\bigg)}{\{({1/m_k})^{m_k}\} \Gamma(m_k)}- \sum_{j=0}^{N_a+l-1}\exp\bigg\{-\bigg(\dfrac{u_4 U_2}{U_1}\bigg)\bigg( {N_a } -\dfrac{U_{1} N_b}{V_{1}}\bigg)\bigg\}  \dfrac{\bigg( {N_a } -\dfrac{U_{1} N_b}{V_{1}}\bigg)^j \sum_{t_l=0}^{j}{{j}\choose{t_l}}\bigg(\dfrac{u_4 U_2}{U_1}\bigg)^{j-t_l} \bigg(\dfrac{u_4}{ U_1}\bigg)^{t_l}}       {j!} $}\\\scalebox{0.8}{$ \int\limits_0^{\infty} \dfrac{\{{z}^{(m_k-r+l-t_l-1)}\} \text{exp}\Bigg\{-{m_k z} -\dfrac{u_4 N_b}{z V_{1}}- \bigg(\dfrac{u_4}{z U_1}\bigg)\bigg( {N_a } -\dfrac{U_{1} N_b}{V_{1}}\bigg)\Bigg\}}{\{({1/m_k})^{m_k}\} \Gamma(m_k)}  dz\Bigg]  + I^{'}$}
\end{multline}

Finally, above equation can be written as follows 
\begin{multline}
\scalebox{0.79}{$\mathscr{P}\bigg(Y_{1}\geq \dfrac{u_4}{Z V_{1}}+\dfrac{u_4 U_2}{V_1}-\dfrac{U_{1} X_{1}}{V_{1}}\bigg) =  \sum_{p_a=0}^{N_b-1} \sum_{r=0}^{p_a}{{p_a}\choose{r}} \sum_{l=0}^{r}(-1)^{l}{{r}\choose{l}} \Big(\dfrac{u_4 U_2 N_b}{V_1}\Big)^{p_a -r}\exp\Big(-\dfrac{u_4 U_2 N_b}{V_1}\Big) \dfrac{\bigg(\dfrac{u_4 N_b}{ V_{1}}\bigg)^{r-l}      \bigg(\dfrac{U_{1}  N_b}{V_{1}}\bigg)^{l} \Gamma(N_a+l)}{p_a!  \big\{({1/N_a})^{N_a}\big\} \bigg( {N_a } -\dfrac{U_{1} N_b}{V_{1}}\bigg)^{N_a+l}  \Gamma(N_a)} $}\\\scalebox{0.775}{$  \Bigg[\dfrac{2{\big(\dfrac{u_4 N_b}{V_1 m_k}\big)}^{\dfrac{m_k+l-r}{2}}K_{m_k+l-r}\bigg(2\sqrt{\dfrac{u_4 {N_b}m_k}{V_1}}\bigg)}{\{({1/m_k})^{m_k}\} \Gamma(m_k)}- \sum_{j=0}^{N_a+l-1}\exp\bigg\{-\bigg(\dfrac{u_4 U_2}{U_1}\bigg)\bigg( {N_a } -\dfrac{U_{1} N_b}{V_{1}}\bigg)\bigg\}  \dfrac{\bigg( {N_a } -\dfrac{U_{1} N_b}{V_{1}}\bigg)^j \sum_{t_l=0}^{j}{{j}\choose{t_l}}\bigg(\dfrac{u_4 U_2}{U_1}\bigg)^{j-t_l} \bigg(\dfrac{u_4}{ U_1}\bigg)^{t_l}} {j! \{({1/m_k})^{m_k}\} \Gamma(m_k)} $}\\\scalebox{0.79}{$           
2{\big(\dfrac{u_4 N_a}{U_1 m_k}\big)}^{\dfrac{m_k+l-r-t_l}{2}}K_{m_k+l-r-t_l}\bigg(2\sqrt{\dfrac{u_4 {N_a}m_k}{U_1}}\bigg)\Bigg]   $}\\\scalebox{0.79}{$+\sum_{p_a=0}^{N_a-1}\dfrac{\bigg({\dfrac{u_{4} {N_{a}}}{U_{1}}}\bigg)^{p_a}  ({m_k})^{m_k} \exp\Big(-\dfrac{u_4 U_2 N_a}{U_1}\Big) \sum_{r=0}^{p_a}{{p_a}\choose{r}} {(U_2)}^{p_a-r} }{p_a! \Gamma(m_k) }2  {\Bigg(      {\dfrac{u_{4} N_a}{U_{1}m_k}}\Bigg)}^{(m_k -r)/2}K_{m_k - r}\bigg(2\sqrt{{{\dfrac{u_{4} {N_{a}}m_k}{U_{1} }}}}\bigg)$}
\end{multline}

I$^{'}$ can be determined as follows [21, Sec. 3.471.9, 8.352.2].

\begin{multline}
\scalebox{0.84}{$I^{'}=\int\limits_0^{\infty} \frac{\{{z}^{(m_k-1)}\} \text{exp}(-\frac{z} {{1/m_k}})}{\{({1/m_k})^{m_k}\} \Gamma(m_k)}\bigg[ \int\limits_{\frac{u_4}{z U_1}+\frac{u_4 U_2}{U_1}}^{\infty} \frac{\{{x_{1}}^{(N_a-1)}\} \text{exp}(-\frac{x_{1}} {{1/N_a}})}{\{({1/N_a})^{N_a}\} \Gamma(N_a)} dx_{1}\bigg]  dz=\dfrac{1}{\Gamma{(N_a)}}   \int\limits_0^{\infty} \frac{\{{z}^{(m_k-1)}\} \text{exp}(-\frac{z} {{1/m_k}})}{\{({1/m_k})^{m_k}\} \Gamma(m_k)}{\Gamma\Bigg(N_a,\dfrac{u_4 N_a}{z U_1}+\dfrac{u_4 U_2 N_a}{U_1}\Bigg)}dz$}\\\scalebox{0.86}{$ =\int\limits_0^{\infty} \frac{\{{z}^{(m_k-1)}\} \text{exp}(-\frac{z} {{1/m_k}})}{\{({1/m_k})^{m_k}\} \Gamma(m_k)}\Bigg[\exp\Big(-\dfrac{u_4 N_a}{z U_1}-\dfrac{u_4 U_2 N_a}{U_1}\Big)\sum_{p_a=0}^{N_a-1}\dfrac{\bigg(\dfrac{u_4 N_a}{z U_1}+\dfrac{u_4 U_2 N_a}{U_1}\bigg)^{p_a}}{p_a!}\Bigg]dz$}\\\scalebox{0.85}{$=\sum_{p_a=0}^{N_a-1}\dfrac{\bigg({\dfrac{u_{4} {N_{a}}}{U_{1}}}\bigg)^{p_a}  ({m_k})^{m_k} \exp\Big(-\dfrac{u_4 U_2 N_a}{U_1}\Big) \sum_{r=0}^{p_a}{{p_a}\choose{r}} {(U_2)}^{p_a-r} }{p_a! \Gamma(m_k) }\int\limits_0^{\infty} {\big(z\big)}^{m_k - r -1} \exp\Bigg(-{m_k z} -\dfrac{u_4 N_a}{z U_1}\Bigg)dz$}\\\scalebox{0.85}{$=\sum_{p_a=0}^{N_a-1}\dfrac{\bigg({\dfrac{u_{4} {N_{a}}}{U_{1}}}\bigg)^{p_a}  ({m_k})^{m_k} \exp\Big(-\dfrac{u_4 U_2 N_a}{U_1}\Big) \sum_{r=0}^{p_a}{{p_a}\choose{r}} {(U_2)}^{p_a-r} }{p_a! \Gamma(m_k) }2  {\Bigg(      {\dfrac{u_{4} N_a}{U_{1}m_k}}\Bigg)}^{(m_k -r)/2}K_{m_k - r}\bigg(2\sqrt{{{\dfrac{u_{4} {N_{a}}m_k}{U_{1} }}}}\bigg)$}
\end{multline}

\subsubsection{Appendix F: Closed form PU outage expression for DF relaying}
\begin{multline}
\mathscr{P}_{out}^{(PU,DF)} = 1 - \bigg[\mathscr{P}\lbrace Q_1 \rbrace\times  \mathscr{P}\big\lbrace R_{PU_1}^{(2,DF)}\geq R_{PU}\big\rbrace \times \mathscr{P}\big\lbrace R_{PU_2}^{(2,DF)}\geq R_{PU}\big\rbrace\bigg]\\ =1-\Bigg[\scalebox{0.9}{$\dfrac{1}{\Gamma(N_a)\Gamma(N_b)}{\Gamma\Bigg(N_a,{{N_a u_1}/{A_1}}\Bigg)\Gamma\Bigg(N_b,{{N_b u_2}/{A_2}}\Bigg)}-\Bigg\{\dfrac{1}{\Gamma({N_a})}-\dfrac{\gamma\Big(N_b,N_b \dfrac{u_{2}}{A_2}\Big)}{\Gamma(N_a) \Gamma(N_b)}\Bigg\}\Big\{\gamma\big(N_a,N_a\dfrac{u_{3}-u_{1}}{A_1}\big)-\gamma\big(N_a,N_a\dfrac{u_1}{A_1}\big)\Big\}$}\\\scalebox{0.85}{$+\sum_{{q_a}=0}^{N_b-1}\dfrac{\exp\big(-\frac{N_b {u_{3}}}{A_2}\big)\sum_{q=0}^{{q_a}}(-1)^{q}{{{q_a}}\choose{q}} {u_3}^{{q_a}-q} {A_1}^q}{{q_a}!  \Gamma(N_a) \big(1/N_a\big)^{N_a} \big(1/N_b\big)^{{q_a}} \Big(N_a-\frac{A_1 N_b}{A_2}\Big)^{(N_a+q)}{A_2}^{q_a}}\Big\{\gamma\big[N_a+q, \big(N_a-\frac{A_1 N_b}{A_2}\big)\dfrac{u_{3}-u_{1}}{A_1}\big]-\gamma\big[N_a+q, \big(N_a-\frac{A_1 N_b}{A_2}\big)\dfrac{u_{1}}{A_1}\big]\Big\}\Bigg] $}\\ \times \Bigg[\dfrac{1}{\Gamma(N_a)}{\gamma\Bigg(N_a,\dfrac{\sqrt{\dfrac{k^{'} }{a_{1}}}}{1/N_a}\Bigg)}-\sum_{p_a=0}^{N_b-1}\dfrac{{N_a}^{N_a} {N_b}^{p_a}}{p_a!  \Gamma(N_a)}\sum_{r=0}^{p_a}(-1)^{r}{{p_a}\choose{r}} \bigg(\dfrac{k^{'}}{b_{1}}\bigg)^{p_a-r} \bigg(\dfrac{a_{1}}{b_{1}}\bigg)^{r} \sum_{l=0}^{\infty} \\ \sum_{t_l=0}^{l} {{l}\choose{t_l}} \dfrac{(-1)^l}{l! (2r+ N_a -p_a-l+2t_l)}  {\Big(\dfrac{k^{'} N_b}{b_{1}}\Big)}^{l-t_l} \Bigg({N_a-\dfrac{a_{1}N_b}{b_{1}}} \Bigg)^{t_l}  \Bigg(\sqrt{\dfrac{k^{'}}{a_{1}}}\Bigg)^{2r+ N_a -p_a-l+2t_l} \Bigg]\\ \times \Bigg[\dfrac{1}{\Gamma(N_b)}{\gamma\Bigg(N_b,\dfrac{\sqrt{\dfrac{k^{''} }{b_{1}}}}{1/N_b}\Bigg)}-\sum_{p_a=0}^{N_a-1}\dfrac{{N_b}^{N_b}{N_a}^{p_a}}{p_a!  \Gamma(N_b)}\sum_{r=0}^{p_a}(-1)^{r}{{p_a}\choose{r}} \bigg(\dfrac{k^{''}}{a_{1}}\bigg)^{p_a-r} \bigg(\dfrac{b_{1}}{a_{1}}\bigg)^{r} \sum_{l=0}^{\infty} \\\sum_{t_l=0}^{l} {{l}\choose{t_l}} \dfrac{(-1)^l}{l! (2r+ N_b -p_a-l+2t_l)}  {\Big(\dfrac{k^{''} N_a}{a_{1}}\Big)}^{l-t_l} \Bigg({N_b-\dfrac{b_{1}N_a}{a_{1}}}\Bigg)^{t_l}  \Bigg(\sqrt{\dfrac{k^{''}}{b_{1}}}\Bigg)^{2r+ N_b -p_a-l+2t_l} \Bigg]
\end{multline}

\subsubsection{Appendix G: Closed form SU outage expression for DF relaying}
\begin{multline}
\mathscr{P}_{out}^{(SU,DF)} = 1 - \big[\mathscr{P}\lbrace Q_1  \rbrace \times \mathscr{P}\lbrace  Q_2 \rbrace \times \mathscr{P}\big\lbrace R_{SU_2}^{(2,DF)}\geq R_{SU}\big\rbrace\big]\\=1-\scalebox{0.92}{$\Bigg[\dfrac{1}{\Gamma(N_a)\Gamma(N_b)}{\Gamma\Bigg(N_a,{{N_a u_1}/{A_1}}\Bigg)\Gamma\Bigg(N_b,{{N_b u_2}/{A_2}}\Bigg)}-\Bigg\{\dfrac{1}{\Gamma({N_a})}-\dfrac{\gamma\Big(N_b,N_b \dfrac{u_{2}}{A_2}\Big)}{\Gamma(N_a)\Gamma(N_b)}\Bigg\}$}\\\scalebox{0.85}{$\Big\{\gamma\big(N_a,N_a\dfrac{u_{3}-u_{1}}{A_1}\big)-\gamma\big(N_a,N_a\dfrac{u_1}{A_1}\big)\Big\}+\sum_{{q_a}=0}^{N_b-1}\frac{\exp\big(-\frac{N_b {u_{3}}}{A_2}\big)\sum_{q=0}^{{q_a}}(-1)^{q}{{{q_a}}\choose{q}} {u_3}^{{q_a}-q} {A_1}^q}{{q_a}!  \Gamma(N_a) (1/N_a)^{N_a} \big(1/N_b\big)^{{q_a}} \Big(N_a-\frac{A_1 N_b}{A_2}\Big)^{(N_a+q)}{A_2}^{q_a}}$}\\\scalebox{0.9}{$\Big\{\gamma\big[N_a+q, \big(N_a-\frac{A_1 N_b}{A_2}\big)\dfrac{u_{3}-u_{1}}{A_1}\big]-\gamma\big[N_a+q, \big(N_a-\frac{A_1 N_b}{A_2}\big)\dfrac{u_{1}}{A_1}\big]\Big\}\Bigg]$}\\\times \scalebox{0.85}{$\Bigg[\dfrac{1}{\Gamma(N_a)\Gamma(N_b)}{\Gamma\Bigg(N_a,{{N_a u_1}/{B_1}}\Bigg)\Gamma\Bigg(N_b,{{N_b u_2}/{B_2}}\Bigg)}-\Bigg\{\dfrac{1}{\Gamma(N_a)}-\dfrac{\gamma\Big(N_b,N_b \dfrac{u_{2}}{B_2}\Big)}{\Gamma(N_a)\Gamma(N_b)}\Bigg\} $}\\\scalebox{0.9}{$ \Big\{\gamma\big(N_a,N_a\dfrac{u_{3}-u_{1}}{B_1}\big)-\gamma\big(N_a,N_a\dfrac{u_1}{B_1}\big)\Big\}+\sum_{{q_a}=0}^{N_b-1}\frac{\exp\big(-\frac{N_b {u_{3}}}{B_2}\big)\sum_{q=0}^{{q_a}}(-1)^{q}{{{q_a}}\choose{q}} {u_3}^{{q_a}-q} {B_1}^q}{{q_a}!  \Gamma(N_a) \big(1/N_a\big)^{N_a} \big(1/N_b\big)^{{q_a}} \Big(N_a-\frac{B_1 N_b}{B_2}\Big)^{(N_a+q)}{B_2}^{q_a}}$}\\\scalebox{0.85}{$\Big\{\gamma\big[N_a+q, \big(N_a-\frac{B_1 N_b}{B_2}\big)\dfrac{u_{3}-u_{1}}{B_1}\big]-\gamma\big[N_a+q, \big(N_a-\frac{B_1 N_b}{B_2}\big)\dfrac{u_{1}}{B_1}\big]\Big\}\Bigg]$}\\\times \scalebox{0.85}{$\Bigg[\sum_{p_a=0}^{N_b-1}  \dfrac{\Big(\dfrac{N_b}{b_1}\Big)^{p_a}}{p_a!  ({1/N_a})^{N_a} \Gamma(N_a)} \sum_{r=0}^{p_a}(-1)^{r}{{p_a}\choose{r}}  \bigg(\dfrac{u_4}{c}\bigg)^{p_a-r} \big({a_{1}}\big)^{r}\frac{1}{{\bigg(N_a-\frac{a_{1}N_b}{b_{1}}\bigg)}^{N_a+r}}\dfrac{\Gamma{(N_a +r)}}{\{({1/m_k})^{m_k}\} \Gamma(m_k)}\bigg[2\bigg\{\frac{u_4 N_b}{c b_{1} m_k} \bigg\}^{(m_k-p_a+r)/2} $}\\\scalebox{0.85}{$ K_{m_k-p_a+r}\bigg\{2\sqrt{\frac{u_4 {N_b} m_k}{c b_{1}}}\bigg\}   - \sum_{j=0}^{N_a+r-1} \dfrac{1}{j!} \bigg[\frac{u_4  \bigg({N_a}-\dfrac{a_{1}{N_b}}{b_{1}}\bigg)}{c a_{1}} \bigg]^{j}  2\bigg\{\frac{u_4  \bigg(N_a-\frac{a_{1}N_b}{b_{1}}\bigg)}{c a_{1} m_k}+\frac{u_4 N_b}{c b_{1} m_k}\bigg\}^{\frac{(m_k-p_a+r-j)}{2}}$}\\\scalebox{0.85}{$ K_{m_k-p_a+r-j}\bigg\{2\sqrt{\frac{u_4  m_k \bigg({N_a}-\frac{a_{1}{N_b}}{b_{1}}\bigg)}{c a_{1}}+\frac{u_4 {N_b} m_k}{c b_{1} }}\bigg\}  \Bigg]+\sum_{p_a=0}^{N_a-1}\dfrac{\bigg({\dfrac{u_{4} {N_{a}}}{c a_{1}}}\bigg)^{p_a}  ({m_k})^{m_k}}{p_a! \Gamma(m_k)} 2  {\Bigg(      {\dfrac{u_{4} N_a}{c a_{1} m_k}}\Bigg)}^{(m_k - p_a)/2}K_{m_k - p_a}\bigg(2\sqrt{{{\dfrac{u_{4} {N_{a}} m_k}{c a_{1} }}}}\bigg)\Bigg]$}
\end{multline}

\subsubsection{Appendix H: Closed form PU outage expression for AF relaying}
\begin{multline}
\mathscr{P}_{out}^{(PU,AF)} = 1 - \bigg[\mathscr{P}\big\lbrace R_{PU_1}^{(2,AF)}\geq R_{PU}\big\rbrace \times \mathscr{P}\big\lbrace R_{PU_2}^{(2,AF)}\geq R_{PU}\big\rbrace\bigg]\\=1- \Bigg[\sum_{{q_a}=0}^{N_b-1}(N_b u_1)^{q_a}\dfrac{\exp\big(-\frac{N_b F_1{u_{1}}}{(C_1-E_1 u_1)}\big)\sum_{q=0}^{{q_a}}{{{q_a}}\choose{q}} \sum_{l=0}^{q}{{q}\choose{l}} {F_1}^{{q_a}-q} {H_1}^{q-l}}{{q_a}!  \Gamma(N_a) \big(1/N_a\big)^{N_a} {(C_1-E_1 u_1)}^{q_a}}\\  2  {\Bigg( {\dfrac{N_b u_1/(C_1-E_1 u_1)}{N_a+\frac{N_b H_1 u_1}{(C_1-E_1 u_1)}}}\Bigg)}^{(N_a+q-2l)/2}K_{N_a+q-2l}\bigg(2\sqrt{\dfrac{N_b u_1}{(C_1-E_1 u_1)}\big({N_a+\frac{N_b H_1 u_1}{C_1-E_1 u_1}}}\big)\bigg)\\ \times \sum_{{q_a}=0}^{N_a-1}(N_a u_1)^{q_a}\dfrac{\exp\big(-\frac{N_a F_2{u_{1}}}{(C_2-E_2 u_1)}\big)\sum_{q=0}^{{q_a}}{{{q_a}}\choose{q}} \sum_{l=0}^{q}{{q}\choose{l}} {F_2}^{{q_a}-q} {H_2}^{q-l}}{{q_a}!  \Gamma(N_b) \big(1/N_b\big)^{N_b} {(C_2-E_2 u_1)}^{q_a}}\\   2  {\Bigg( {\dfrac{N_a u_1/(C_2-E_2 u_1)}{N_b+\frac{N_a H_2 u_1}{(C_2-E_2 u_1)}}}\Bigg)}^{(N_b+q-2l)/2}K_{N_b+q-2l}\bigg(2\sqrt{\dfrac{N_a u_1}{(C_2-E_2 u_1)}\big({N_b+\frac{N_a H_2 u_1}{C_2-E_2 u_1}}}\big)\bigg) \Bigg]
\end{multline}

\subsubsection{Appendix M: Closed form SU outage expression for AF relaying}
\begin{multline}
\mathscr{P}_{out}^{(SU,AF)} = \scalebox{0.84}{$1- \Bigg[ \sum_{p_a=0}^{N_b-1} \sum_{r=0}^{p_a}{{p_a}\choose{r}} \sum_{l=0}^{r}(-1)^{l}{{r}\choose{l}}\Big(\dfrac{u_4 U_2 N_b}{V_1}\Big)^{p_a -r}\exp\Big(-\dfrac{u_4 U_2 N_b}{V_1}\Big) \dfrac{\bigg(\dfrac{u_4 N_b}{ V_{1}}\bigg)^{r-l}      \bigg(\dfrac{U_{1}  N_b}{V_{1}}\bigg)^{l} \Gamma(N_a+l)}{p_a! \big\{({1/N_a})^{N_a}\big\} \bigg( {N_a } -\dfrac{U_{1} N_b}{V_{1}}\bigg)^{N_a+l}  \Gamma(N_a)}$}\\\scalebox{0.67}{$  \bigg[\dfrac{2{\big(\dfrac{u_4 N_b}{V_1 m_k}\big)}^{\dfrac{m_k+l-r}{2}}K_{m_k+l-r}\bigg(2\sqrt{\dfrac{u_4 {N_b}m_k}{V_1}}\bigg)}{\{({1/m_k})^{m_k}\} \Gamma(m_k)}- \sum_{j=0}^{N_a+l-1}\exp\bigg\{-\bigg(\dfrac{u_4 U_2}{U_1}\bigg)\bigg( {N_a } -\dfrac{U_{1} N_b}{V_{1}}\bigg)\bigg\}   \dfrac{\bigg( {N_a } -\dfrac{U_{1} N_b}{V_{1}}\bigg)^j \sum_{t_l=0}^{j}{{j}\choose{t_l}}\bigg(\dfrac{u_4 U_2}{U_1}\bigg)^{j-t_l} \bigg(\dfrac{u_4}{ U_1}\bigg)^{t_l}}       {j! \{({1/m_k})^{m_k}\} \Gamma(m_k)}           
2{\big(\dfrac{u_4 N_a}{U_1 m_k}\big)}^{\dfrac{m_k+l-r-t_l}{2}}$}\\\scalebox{0.8}{$K_{m_k+l-r-t_l}\bigg(2\sqrt{\dfrac{u_4 {N_a}m_k}{U_1}}\bigg)\bigg]   +\sum_{p_a=0}^{N_a-1}\dfrac{\bigg({\dfrac{u_{4} {N_{a}}}{U_{1}}}\bigg)^{p_a}  ({m_k})^{ m_k} \exp\Big(-\dfrac{u_4 U_2 N_a}{U_1}\Big)}{p_a! \Gamma(m_k) } \sum_{r=0}^{p_a}{{p_a}\choose{r}} {(U_2)}^{p_a-r} 2  {\Bigg(      {\dfrac{u_{4}N_a}{U_{1}m_k}}\Bigg)}^{(m_k -r)/2}K_{m_k - r}\bigg(2\sqrt{{{\dfrac{u_{4} {N_{a}}m_k}{U_{1} }}}}\bigg) \Bigg] $}\\\scalebox{0.82}{$\times \Bigg[ \sum_{p_a=0}^{N_b-1} \sum_{r=0}^{p_a}{{p_a}\choose{r}} \sum_{l=0}^{r}(-1)^{l}{{r}\choose{l}}\Big(\dfrac{u_s U_2 N_b}{S_2}\Big)^{p_a -r}\exp\Big(-\dfrac{u_s U_2 N_b}{S_2}\Big) $}\\\scalebox{0.7}{$ \dfrac{\bigg(\dfrac{u_s N_b}{ S_{2}}\bigg)^{r-l}      \bigg(\dfrac{S_{1}  N_b}{S_{2}}\bigg)^{l} \Gamma(N_a+l)}{p_a! \big\{({1/N_a})^{N_a}\big\} \bigg( {N_a } -\dfrac{S_{1} N_b}{S_{2}}\bigg)^{N_a+l}  \Gamma(N_a)}\bigg[\dfrac{2{\big(\dfrac{u_s N_b}{S_2 m_k}\big)}^{\dfrac{m_k+l-r}{2}}K_{m_k+l-r}\bigg(2\sqrt{\dfrac{u_s {N_b}m_k}{S_2}}\bigg)}{\{({1/m_k})^{m_k}\} \Gamma(m_k)}$}\\\scalebox{0.7}{$  - \sum_{j=0}^{N_a+l-1}\exp\bigg\{-\bigg(\dfrac{u_s U_2}{S_1}\bigg)\bigg( {N_a } -\dfrac{S_{1} N_b}{S_{2}}\bigg)\bigg\} 2{\big(\dfrac{u_s N_a}{S_1 m_k}\big)}^{\dfrac{m_k+l-r-t_l}{2}}  \dfrac{\bigg( {N_a } -\dfrac{S_{1} N_b}{S_{2}}\bigg)^j \sum_{t_l=0}^{j}{{j}\choose{t_l}}\bigg(\dfrac{u_s U_2}{S_1}\bigg)^{j-t_l} \bigg(\dfrac{u_s}{ S_1}\bigg)^{t_l}}       {j! \{({1/m_k})^{m_k}\} \Gamma(m_k)} K_{m_k+l-r-t_l}\bigg(2\sqrt{\dfrac{u_s {N_a}m_k}{S_1}}\bigg)\bigg]           
$}\\\scalebox{0.75}{$  +\sum_{p_a=0}^{N_a-1}\dfrac{\bigg({\dfrac{u_{s} {N_{a}}}{S_{1}}}\bigg)^{p_a}  ({m_k})^{m_k} \exp\Big(-\dfrac{u_s U_2 N_a}{S_1}\Big)}{p_a! \Gamma(m_k) } \sum_{r=0}^{p_a}{{p_a}\choose{r}} {(U_2)}^{p_a-r} 2  {\Bigg(      {\dfrac{u_{s} N_a}{S_{1} m_k}}\Bigg)}^{(m_k -r)/2}K_{m_k - r}\bigg(2\sqrt{{{\dfrac{u_{s} {N_{a}}m_k}{S_{1} }}}}\bigg) \Bigg]$}
\end{multline}

\bibliographystyle{IEEEtran}
\bibliography{IEEEabrv,library}

\end{document}